\documentclass[review]{elsarticle}
\graphicspath{ {./figures/} }
\usepackage{hyperref}
\usepackage{float}
\usepackage{verbatim} %comments
\usepackage{apalike}
\restylefloat{figure}
\floatstyle{plaintop} %table caption at top
\restylefloat{table}
\usepackage[utf8]{inputenc} 
\usepackage[T1]{fontenc}
\usepackage[hungarian,english]{babel}

\newenvironment{abstracts}
 {\global\setbox\absbox=\vbox\bgroup
    \hsize=\textwidth
    \linespread{1}\selectfont}
 {\vspace{-\bigskipamount}\egroup}
\renewenvironment{abstract}[1][]
 {\if\relax\detokenize{#1}\relax\else\selectlanguage{#1}\fi
  \noindent\textbf{\abstractname}\par\medskip\noindent\ignorespaces}
 {\par\bigskip}

\journal{Magyar Tudomány}

\biboptions{authoryear}

\begin{document}
\begin{frontmatter}

%The title page must contain the title of the paper and the full name/full affiliation with country/e-mail address for each author and co-author of the manuscript. Please make sure you have included all elements listed below with your manuscript submission.

\title{Mesterséges Intelligencia Kutatások Magyarországon}

\author[label1,label2]{Benczúr András}
\ead{benczur@sztaki.hun-ren.hu}

\author[label3]{Gyimóthy Tibor}
\ead{gyimothy@inf.u-szeged.hu}

\author[label4]{Szegedy Balázs}%  \corref{cor1}}
\ead{szegedy@renyi.hu}

%\cortext[cor1]{Corresponding author.}
\address[label1]{HUN-REN SZTAKI}
\address[label2]{Széchenyi Egyetem, Győr}
\address[label3]{Szegedi Tudományegyetem}
\address[label4]{HUN-REN Rényi Alfréd Matematikai Kutatóintézet}

\begin{abstracts}

\begin{abstract}[hungarian]
    A mesterséges intelligencia (MI) a 2000-es évek közepétől rendkívül látványos fejlődésnek indult, különösen a gépi tanulás és azon belül a mélytanulás (deep learning) területén, valamint a nagy adatbázisok és a számítási kapacitás robbanásszerű növekedésének köszönhetően. A hazai kutatók korán felismerték az MI jelentőségét, bekapcsolódtak a nemzetközi kutatásokba, és jelentős eredményeket értek el mind elméleti, mind gyakorlati területeken. A cikk a hazai MI-kutatások néhány kiemelkedő eredményét ismerteti. Bemutatja a mélytanulás berobbanása előtti időszak (a 2010-es évek eleje) eredményeit, majd ismerteti a 2010 utáni jelentősebb hazai elméleti kutatásokat. Végezetül rövid áttekintést ad a 2010 utáni, MI-hez kapcsolódó alkalmazásorientált tudományos eredményekről. 

\noindent 
\textit{Kulcsszavak:}
    Mesterséges intelligencia; mélytanulás; hazai kutatások; alkalmazások 
\end{abstract}
\begin{abstract}
Artificial intelligence (AI) has undergone remarkable development since the mid-2000s, particularly in the fields of machine learning and deep learning, driven by the explosive growth of large databases and computational capacity. Hungarian researchers recognized the significance of AI early on, actively participating in international research and achieving significant results in both theoretical and practical domains.  This article presents some key achievements in Hungarian AI research. It highlights the results from the period before the rise of deep learning (the early 2010s), then discusses major theoretical advancements in Hungary after 2010. Finally, it provides a brief overview of AI-related applied scientific achievements from 2010 onward. 

\end{abstract}
\end{abstracts}

\end{frontmatter}

\section{Bevezetés}

A mesterséges intelligencia a 2000-es évek közepétől rendkívül látványos fejlődésnek indult a gépi tanulás, különösen a mélytanulás (deep learning), valamint a nagy adatbázisok és a számítási kapacitás robbanásszerű növekedésével. Az interneten keresztül hatalmas mennyiségű adat állt rendelkezésre, és a grafikus processzorok (GPU-k) elősegítették a komplex neurális hálózatok kiképzését olyan kiemelkedő jelentőségű alkalmazási területeken, mint a  képfelismerés, természetes nyelvfeldolgozás és önvezető autók.

A mesterséges intelligencia számos hullámon keresztül jutott el a mai szintre, amelyet technológiai előrelépések, kutatási eredmények, finanszírozási ciklusok és a mesterséges intelligenciával kapcsolatos társadalmi elvárások határoztak meg. Az 1950-es években megszületett a Turing-teszt koncepciója, az első egyszerű gépi tanulás kísérletek alapján a kutatók úgy hitték, hogy a mesterséges intelligencia néhány évtizeden belül megoldhatja a legtöbb emberi problémát. Az elvárások azonban túlzottak voltak, az 1970-es évek közepén az optimizmus kifulladt. Az 1980-as években az AI új lendületet kapott a szakértői rendszerek megjelenésével. Ezek azonban kudarcot vallottak, mivel statikusak voltak, nem voltak képesek önállóan tanulni vagy alkalmazkodni új helyzetekhez, költségesek és korlátozottan skálázhatók voltak. Két csalódottsági fázis, úgynevezett ,,AI Winter'' után, a mai korszakban az AI dinamikusan fejlődik, de a múlt tanulságai miatt óvatosabb elvárásokkal közelítik meg a technológia lehetőségeit és korlátait.

A hazai kutatók korán felismerték az MI terület jelentőségét. Bekapcsolódtak a nemzetközi kutatásokba és jelentős eredményeket értek el mind elméleti mind pedig a gyakorlati területeken. Ugyanakkor kevés olyan tanulmány született, ami áttekintést adna a hazai MI kutatásokról. Kivételt jelent Sántáné-Tóth Edit tanulmánya \cite{SantaneToth1} illetve Dömölki Bálint a Mesterséges Intelligencia Múltja Magyarországon címmel megjelent írása \cite{Domolki}. Sántáné-Tóth Edit eredetileg az 1996-ban Budapesten megrendezett European Conference on Artificial Intelligenc (ECAI’96) konferencia alkalmából készített egy összefoglalót a hazai MI kutatásokról és ipari fejlesztésekről. A hivatkozott tanulmány ennek bővített változata. A szerző bemutatja, hogy milyen szervezeti keretek között alakult ki a hazai MI kutatás-fejlesztés illetve megemlíti a legfontosabb hazai MI kutatási területeket is a kapcsolódó intézményekkel együtt. A tanulmány kiegészítésként egy bibliográfia is készült a hazai MI kutatásokról \cite{SantaneToth2}.

Dömölki Bálint informatív tanulmányában  a magyar mesterséges intelligencia kutatások néhány fontosabb eredményét emeli ki, amelyek egyrészt a Magyarországon folyó számítástudományi fejlesztések fő irányához tartoztak, másrészt nemzetközi jelentőséggel is rendelkeztek. A  tanulmány hivatkozásai  az NJSZT Informatikatörténeti Adattárában (iTA) összegyűjtött és tárolt eredményekre, tényekre mutatnak, amelyek jól reprezentálják  a magyar informatika történet eseményeit.

Ebben a cikkünkben   hazai MI kutatások ismertetésénél  arra törekedtünk, hogy röviden ugyan, de   konkrét kutatási eredményeket mutassunk be. Olyan eredményeket, amiket a szerzők  az adott terület színvonalas tudományos fórumain publikáltak. 
Magyarországon már a 70-es években is folytak az MI területeihez kapcsolódó kutatások, azonban mint mindenütt a világon nálunk is a mélytanulás előre törésével indult el az MI kutatások robbanásszerű fejlődése. Ezért döntöttünk úgy, hogy cikkünkben három  külön fejezetben tárgyaljuk az MI kutatások bemutatását. A következő  fejezet a mélytanulási eredmények berobbanása előtti időszak (2010-es évek eleje) eredményeit tárgyalja,  majd a 2010 utáni jelentősebb hazai elméleti eredmények bemutatása következik. Végezetül egy rövid áttekintést adunk a 2010 utáni MI-hez kapcsolódó alkalmazási jellegű tudományos eredményekről.  Természetesen ez utóbbi fejezetben a mélytanulási eredmények bemutatása mellett igyekszünk az egyéb jelentős MI eredményeket is ismertetni.
 
\section{Az MI kutatások múltja Magyarországon}
 
\subsection{Kezdetek}
JohnMcCharty amerikai matematikus 1955-ben egy 2 hónapos nyári szeminárium szervezésére tett javaslatot Dartmouthban. A szervezési javaslatban a szeminárium témájaként elsőként használja az ,,Artificial Intelligence'' megnevezést. Sokan ezért ezt a szemináriumot tekintek a MI indulási eseményének. Végül a  szeminárium 1956-ban lett megrendezve, többek közt Marvin Minsky, Claude Shannon és Ray Solomonoff  részvételével. Érdekesség, hogy  a szervezők MI területként az alábbi diszciplínákat jelölték meg: computers, natural language processing, neural networks, theory of computation, abstraction, creativity.

Alig hat évvel a neves esemény után Kalmár László a Szegedi Tudományegyetem (korábban József Attila Tudományegyetem) professzora Colloquium on the Foundations of Mathematics, Mathematical Machines and Their Applications címmel szervezett konferenciát Tihanyba. A rendezvényen a hazai matematikusok mellett többek közt részt vett Alonzo Church, JohnMcCarthy és Yehoshua Bar-Hille is. A konferenciának  Mesterséges intelligencia, gépi tanulás szekciója is volt.

\begin{figure}
\includegraphics{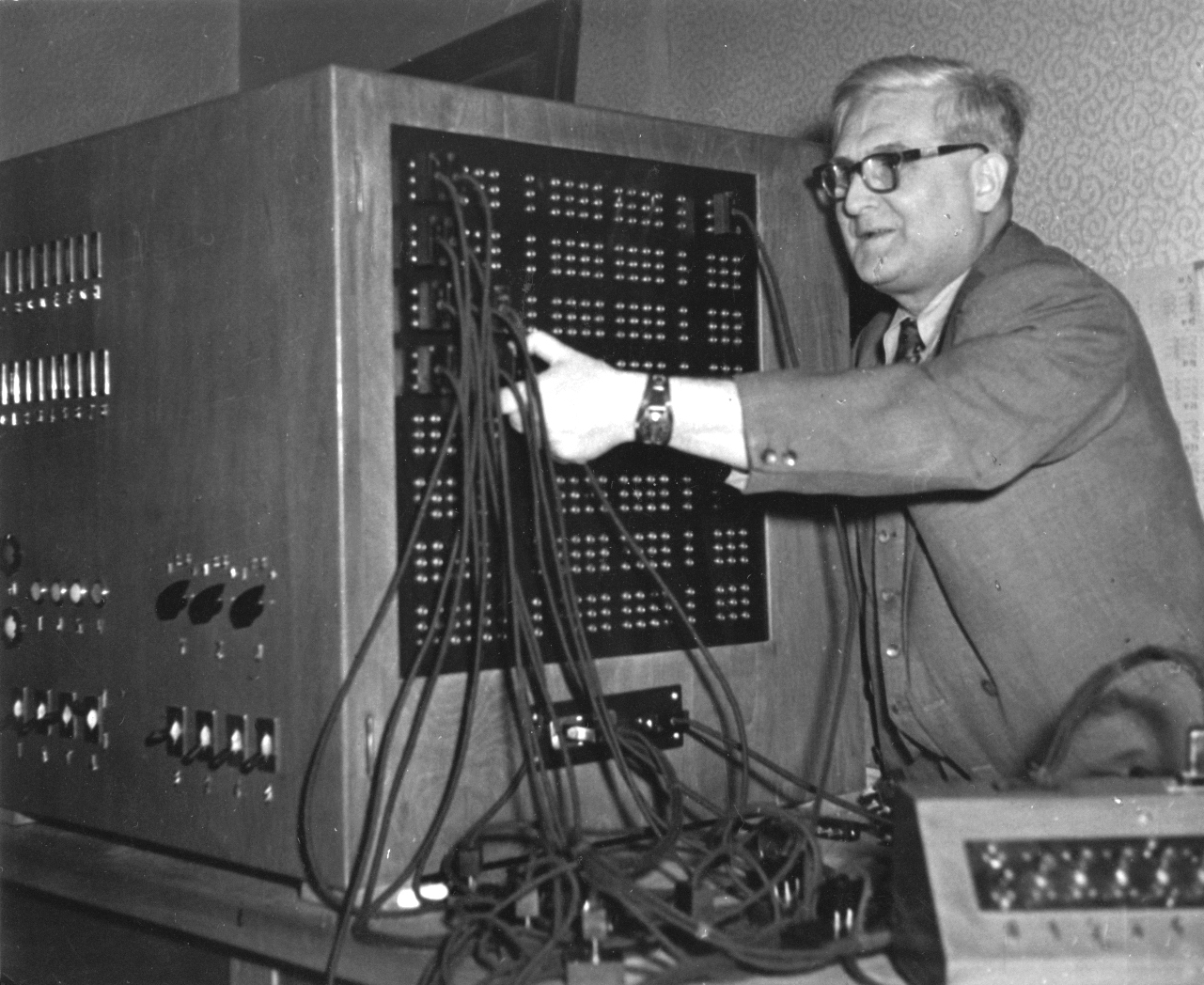}
\caption{Kalmár László}
\end{figure}

Kalmár Lászlót tekinthetjük a hazai MI kutatások elindítójának. A  matematikai logika iránt már az 1920-as években elkezdett érdeklődni és Péter Rózsával együtt kiváló eredményeket értek el ezen a területen. Az elektronikus számítógépek megjelenésével szinte azonnal felismerte a bennük rejlő lehetőségeket. 1956-ban szemináriumot szervezett a szegedi egyetemen a matematikai logika műszaki alkalmazási lehetőségének feltárására. Ezután Kalmár László tervei alapján Muszka Dániel 1958-ban  megépítette a Kalmár féle logikai gépet, amely segítségével az ítéletkalkulus  logikai formuláiról lehetett eldönteni, hogy azok mikor kielégíthetők. A logikai gép ötletéből kiindulva  1959-ben egy varsói konferencián ismertette egy formulavezérlésű számítógép tervét. Az ilyen számítógép anyanyelve nem alacsonyszintű gépi nyelv, hanem egy magasabb szintű programozási nyelv. Kalmár László munkásságáról Szabó Péter Gábor készített  egy kiváló, rövid összefoglalót \cite{Szabo}.
 
\subsection{Elméleti eredmények}
Ebben a fejezetben néhány konkrét eredmény bemutatásával igyekszünk áttekintést adni a hazai 2010 előtti elméleti jellegű MI kutatásokról. A gépi tanulással kapcsolatos eredmények mellett itt szerepelnek az emberi viselkedés és gépi látással kapcsolatos munkák illetve a korai MI kutatásokban jelentős szerepet játszó  Prolog nyelv hazai fejlesztése is említésre kerül.

Szepesvári Csaba a SZTAKI munkatársa kollégájával Kocsis Leventével kidolgoztak egy algoritmust (UCT) a Monte Carlo alapú fa-keresési stratégia optimalizálására \cite{Kocsis}. Ez az algoritmus jobb elméleti alapokkal rendelkezik mint a korábbi algoritmusok és emellett a gyakorlatban is jobban teljesített azoknál. Ennek köszönhetően a módszer alapvető jelentőségű lett a megerősítéses tanulás területén, az UCT algoritmus nagyon sok kutatás-fejlesztési  munka kiindulási  pontjává vált. Többek között az algoritmus egyik változatát használta a DeepMind MuZero programja. Ez volt  az első olyan go program volt, ami erősebbnek bizonyult a legerősebb emberi játékosnál.  Az eredmény 2016-ban ECML/PKDD Test of Time Award-ban részesült.

Lőrincz András az ELTE Mesterséges Intelligencia tanszék munkatársa az agyi folyamatok tanulmányozására építette kutatásait. James Chrobakkal és Buzsáki Györggyel együttműködve olyan modellt készítettek, ami  megmutatta, hogy a humán tanulás egyik fontos eleme lehet a nagy mennyiségű szenzoros információból rejtett, közel független összetevők és közel önálló folyamatok azonosítása \cite{Chrobak}. Szita Istvánnal publikált cikkükben megmutatták, hogy a  komponenseken alapuló megközelítések a megerősítéses tanulásba (ahol a gépek úgy tanulnak, hogy bár késleltetve, de jutalmat vagy büntetést kapnak a válasz sorozatok eredményeképpen) is belefoglalhatók és jelentősen lerövidíthetik a tanulási időt \cite{Szita}.

Vicsek Tamás az ELTE Biológiai Fizika tanszékének kutatója munkatársaival olyan  emberi viselkedéseket vizsgáltak, amelyek kevés paraméterrel modellezhetők. Ilyen eset például, amikor sok ember pánikszerűen menekül egy szobából. Egy 2000-ben a  Nature folyóiratban megjelent cikkükben \cite{Helbing} megmutatták, hogy pánik során sok ember viselkedését (mozgását) 4-5 paramétert tartalmazó egyenletrendszerrel valósághűen lehet a számítógéppel szimulálni, és az eredményekből hasznos következtetések vonhatók le.

SZTAKI-ban  a Roska Tamás vezette kutatócsoportnak  nemzetközileg is vezető szerepe volt a Celluláris Neurális Hálózat (CNN) kutatásában \cite{Roska,Chua}. A matematikai háttér, fizikai modellek  valamint a biológiai analógiák közel hozták az alapszintű mesterséges látás és a biológiai modellek kapcsolatát. Erre a célra speciális analóg három- ill. 4-rétegű neurális hálókon alapuló chipek is készültek  16x16 ill. jobb felbontással. Az új diszciplína ígéretesnek bizonyult, és serkentőleg hatott a gépi látás és az emberi látás modellezésében, és a köztük lévő kapcsolat kutatásában. Ez a kutatás elhalványult, amikor megjelentek a mélytanuló hálózatok és az azokat támogató nagy-összekötöttségű (D-CNN), részben a grafikus processzorok  technológiájára épülő chipek. A ,,régi'' CNN technológia egyes eredményei pedig alkalmazásba kerültek speciális képfeldolgozó hardverekben, textúra detekcióban, gyors felismerő rendszerekben \cite{Sziranyi}.

Csuhaj Varjú Erzsébet az ELTE Algoritmusok és Alkalmazásaik tanszék kutatója egyik elindítója és kiépítője volt a grammatika rendszerek elméletének, amely alkalmassá tette a formális nyelvek és automaták eszköztárát a multi-ágens rendszerek szintaktikai modellezésére \cite{CsuhajVarju1}.  A tárgykört és annak egyik részterületét elindító,  Jürgen Dassow szerzőtársával írt dolgozatában definiálta a tábla architektúra alapú problémamegoldó rendszer formális nyelvi modelljét kooperáló osztott grammatika rendszer néven, ahol a problémamegoldó ágenseket együttműködő formális nyelvtanok, a probléma megoldását formális nyelvek reprezentálták \cite{CsuhajVarju2}.

Szeredi Péter az SZKI munkatársa majd pedig később  a BME kutatója  1975-ben a világon másodikként készített el Prolog interpretert, aminek köszönhetően számos kísérleti Prolog alkalmazás született Magyarországon az 1970-80-as években. 1978-1987 között az MPROLOG rendszer fejlesztésének volt a szakmai vezetője -- ez az egyik első olyan magyar szoftver termék volt, amelyet főleg külföldön forgalmaztak. 1987 és 1990 között az Egyesült Királyságban dolgozott David H.D. Warren csoportjában, ahol multiprocesszoros Prolog implementációk fejlesztésével foglalkozott \cite{Calderwood}.

Az MTA-SZTE Mesterséges Intelligencia Kutatócsoport (RGAI) 1996-ban alakult meg Szegeden  Csirik János vezetésével. Kutatásaiknak jelentős  lendületet adott, hogy a 90-es években csatlakozhattak egy nagy nemzetközi gépi tanulási kutatási projekthez  (Inductive Logic Programming).

Turán György az RGAI  kutatócsoport munkatársa  számitógépes tanulás elmélettel és kapcsolódó bonyolultságelméleti kérdésekkel foglalkozott. A neuronhálózatok diszkrét változatára, a küszöbfüggvény hálózatokra Hajnal Andrással, Wolfgang Maass-szal, Pavel Pudlákkal és Szegedy Márióval exponenciális alsó becslést adtak  2-mélységű, polinomiális súlyú hálózatok bonyolultságára \cite{Hajnal}. Ennek az eredménynek a kiterjesztése tetszőleges súlyokra, illetve legalább 3-mélységű hálózatokra jelenleg is megoldatlan probléma. Számos interaktiv (kérdésekkel való) tanulási modellt vizsgáltak, például Dana Angluinnal, Martins Krikis-szel és Robert Sloannal arra az esetre adtak tanuló algoritmusokat, ahol a kérdésekre adott válaszok hiányosak vagy hibásak is lehetnek \cite{Angluin}. 

Jelasity Márk az RGAI majd  Szegedi Tudományegyetem  dolgozójaként  a sok lokális optimummal rendelkező függvények optimalizálása területén ért el jelentős eredményt. Szerzőtársával olyan genetikus algoritmust javasolt, amely a fajok keletkezésének modellezése alapján több, hierarchikusan strukturált lokális populációt hozott létre dinamikusan, amelyek egy-egy lokális optimumba konvergáltak \cite{Jelasity}. Későbbi munkájában kollégáival pletyka algoritmusokon dolgoztak, ahol nagy számú ágens véletlenszerű kommunikációja segítségével valósítottak meg gépi tanulási algoritmusokat \cite{Ormandi}. 

Dombi József  az RGAI majd a Szegedi Tudományegyetem munkatársaként  folytonos logikák műveleteinek kutatása területén ért el jelentős eredményeket. Megalkotott egy fuzzy-operátor családot, amelynek komoly nemzetközi visszhangja volt,  több kutató használta munkájában \cite{Dombi1,Dombi2}. 

Gyimóthy Tibor az RGAI kutatójaként azt vizsgálta,hogy a szoftverfejlesztés területén kifejlesztett algoritmikus nyomkövető módszerek hogyan alkalmazhatók logikai programok tanulására. Kidolgoztak egy olyan ,,theory revision'' eljárást, amely negatív példák felhasználásával bizonyos esetekre egy általános logikai programból képes előállítani egy adott feladat megoldására alkalmas programot \cite{Alexin1}.

A komputeralgebrai, komputerszámelméleti kutatások az 1980-as években kezdődtek a Kossuth Lajos Tudományegyetemen, Debrecenben. Bár ezeket nem szokás a mesterséges intelligenciába sorolni, a formális matematikai átalakítások végrehajtása komoly intellektuális eredmény és ebben a számítógép ma már sokkal eredményesebb, mint a gyakorló matematikus. Pethő Attila diofantikus egyenletek megoldására dolgozott ki algoritmusokat, legjelentősebb eredménye az elliptikus görbékre illeszkedő egész \cite{Gebel} és S-egész pontok \cite{Pethő} meghatározására vonatkozik.

Gergely Tamás az Alkalmazott Logikai Laboratórium munkatársa kiemelkedő eredményeket  ért el az elméleti számítástudomány megalapozása területén \cite{Gergely,Andréka}. Egy matematikai alapokra épülő egységes számítástudomány létrehozásával foglalkozott, kidolgozva egy konstruktív programozáselméletet hatékony specifikációs eszközökkel és módszertannal. Eszközt adott programozási logikák egységes logikai megalapozására; a szoftver engineering területén kidolgozott egy egységes konstruktív logikai keretet a programok jelentésének egzakt meghatározására és verifikálására.

Vámos Tibor, aki a SZTAKI alapítója volt meghatározó szerepet töltött be a hazai mesterséges intelligencia kutatások kialakulásában. Eredményeket ért el a számítógépes alakfelismerés \cite{Vámos1}   illetve robotika \cite{Siegler} területén. Fő kutatási területe az episztemológia, azaz a megismerés tudománya volt, amely a valóság és az agyunkban róla keletkező kép elválasztásának kérdéseit vizsgálja \cite{Vámos2}.

Kóczy László a BME TMIT kutatója az 1970-es évek első felében kezdett fuzzy témában publikálni és hamarosan kapcsolatba került az akkoriban a világ élvonalába tartozó fuzzy kutatókkal. A 80-as évektől kezdődően számos publikációja jelent meg Hirota Kaoruval, először a digitalis elemi tárolók fuzzy általánosításairól \cite{Koczy1}, majd teljesen új ötletként, a fuzzy szabályalapú rendszerekben alkalmazható interpolációs algoritmusokról \cite{Koczy2}. 

\subsection{Alkalmazások}

Az MI technológiában rejlő alkalmazási lehetőségek  nagyon korán felkeltették a hazai kutató közösségek érdeklődést. Számos nemzetközileg is  jelentős eredmény született a beszédtechnológia, természetes nyelvi feldolgozás, alakfelismerés, gyártástudomány, intelligens mérőműszerek, robotika területén. Néhányat ismertetünk  ezekből az eredményekből.
A BME Távközlési és Médiainformatikai tanszéken Gordos Géza vezetésével már a 70-es években létrejött egy  beszédkutatási csapat, ami azóta is eredményesen működik ezen a területen. Fő eredményük a  magyar nyelvű gépi beszédkeltés és beszédfelismerés technológiai több generációjának kidolgozása és gyakorlati alkalmazások megvalósítása \cite{Nemeth}. Legfontosabb korai alkalmazásaik a MultiVox formáns szintézisen alapuló szövegfelolvasó család 12 nyelven illetve a  ProfiVox-Diád elemösszefűzésen alapuló gépi szövegfelolvasó rendszer változatai \cite{Olaszy}. Az utóbbit integrálták a Jaws for Windows képernyő felolvasó programba, ami a mai napig a legelterjedtebben használt program hazánkban látássérült emberek számára. A ProfiVox-Korpusz elem kiválasztásos technológia működik a MÁV hangos utastájékoztatóban számos állomáson.

A SZTAKI munkatársai korán felismerték, hogy a mesterséges intelligencia alkalmazása milyen lehetőségeket jelent a gyártástudomány és -technológia területén. ,,Efficient use of deficient knowledge'' írta Hatvany József, aki a sokszínű, töredékes, akár ellentmondásos mérnöki tudás megragadásának és hasznosításának eszköztárát látta meg benne, amit Márkus Andrással együtt majd’ négy évtizede szisztematikusan fel is mértek \cite{Markus}. Hazai kutatók nevéhez fűződik a genetikus algoritmusok egyik első műszaki alkalmazása, amiben egy ma ontológiának nevezett módszerrel írták le a technológia tervezés fogalmi rendszerét \cite{Vancza}. Hasonlóan, SZTAKI  kutatója nemzetközi szinten  elsők közt javasolta mesterséges neurális hálózatok alkalmazását megmunkálási folyamatok modellezésére és felügyeletére \cite{Monostori1}. Monostori László 1991-ben megszervezte Budapesten az első ,,IMS '91 — Learning in Intelligent Manufacturing Systems'' nemzetközi konferenciát, majd munkatársaival összegezte, hogy az akkori gépi tanulási módszerek hol és miként szolgálhatják a gyártással kapcsolatos műszaki problémák megoldását \cite{Monostori2}. A későbbiekben úttörő szerepe volt az ágens-alapú módszerek bevezetésében \cite{Monostori3,Csaji}.

A BME Méréstechnika és Információs Rendszerek tanszékének dolgozói már a 80-as évek közepétől intenzív kutatás-fejlesztési tevékenységet végeztek az intelligens mérőrendszerek területén \cite{Papp1,Sztipanovits}. Megalkották a valós-idejű alkalmazásokhoz illeszkedő LISP nyelvet  és ezen a bázison a valós-idejű szakértő keretrendszert, amit sikerrel használtak számos ipari alkalmazásban \cite{Papp2}. Kutatásaik természetes módon egyre szélesebb területre terjedt ki, például több ágenses rendszerekre \cite{Meszaros}, ember-gép kapcsolatra, továbbá a tudásábrázolás és tudásfúzió különböző formáira \cite{VarkonyiKoczy}. A 2000-es évektől Horváth Gábor vezetésével megerősödtek a tanszéken a neurális kutatások. Komplex fizikai rendszerek modellezésére \cite{Valyon} és az azokra alapozó orvosi döntéstámogatás képi diagnosztikájára születtek eredmények \cite{Horvath}. Ezzel párhuzamosan foglalkoztak ontológia alapú természetes nyelvi elemzési problémákkal is \cite{Varga}.

A BME Automatizálási és Alkalmazott Informatika tanszék kutatói a modell alapú kód generálás, valamint a domain specifikus modellezéssel és metamodellezéssel kapcsolatos területeken értek el eredményeket \cite{Lengyel,Forstner1}. A technológia lehetővé tette, hogy komplex szoftveres feladatokat egyszerűsített nyelven és metamodellek szintjén lehessen definiálni és a komplex kód generálást az intelligens algoritmusok képesek voltak megoldani a betanított minták alapján. Peer-to-peer intelligens algoritmusokat fejlesztettek hálózati protokollok megvalósíthatóságára  mobil környezetben \cite{Forstner2,Ekler}. A tanszék jellegzetessége, hogy az informatika mellett villamosmérnöki területekkel is foglalkozik, ahol többek között intelligens robot vezérlés is a kutatási területek közé tartozik. Intelligens autonóm robot kutatásában  vett részt a kutató csoport, többek között automata navigációs megoldások fejlesztése és újrahasznosító robotok fejlesztése területén  értek el eredményeket \cite{Kiss1,Kiss2}.

A MorphoLogic Kft-ben Prószéky Gábor vezetésével  alapvető eredmények  születtek a hazai nyelvtechnológia területén. A legelső eredmény a magyar nyelvhez 1991-ben elkészített Humor (High-speed Unification Morphology) morfológiai elemző program és a hozzá tartozó leíró formanyelv volt \cite{Prószéky1}. Kidolgozták a  ,,Helyes-e?'' helyesírás-ellenőrző módszert, ami  a szavak értelmezését  a formális morfológiai elemzés alapján végzi  \cite{Prószéky2}. A választékos fogalmazás támogatására a MorphoLogic Kft.\ munkatársai  létrehoztak egy toldalékoló szinonimaszótárt, a Helyette rendszert \cite{Prószéky3}. Ez három, nyelvi szempontból fontos funkciót valósít meg: felismeri a forrás-szóalak szótári tövét, megkeresi a forrás-szó jelentés köreit, és az azokhoz tartozó szinonimákat; majd visszaírja a szövegbe a kiválasztott szinonima megfelelő alakját. A Helyette megjelenésekor gyakorlatilag az első és egy ideig az egyetlen olyan programrendszer volt a világon, ami ezen a módon működött.

Az MTA Nyelvtudományi Intézetben Váradi Tamás vezetésével kiemelkedő eredmények születtek a magyar nyelvtechnológiai területén. 1997-ben megalapította a Korpusznyelvészeti Osztályt, ahol vezetésével elkészült a Magyar Nemzeti Szövegtár  \cite{Oravecz}. Az intézet munkatársai jelentős eredményeket értek el a gépi fordítás \cite{Senellart}, és párhuzamos-korpusz-építés \cite{Váradi} területén is.

A Debreceni Egyetemen  Papp Ferenc vezetésével már a 60-as évektől folytak kutatások a számítógépes nyelvészet területén. Az amerikai Brown Corpus után elsőként hozta létre  egy nyelv (magyar)   számítógépes szótárát \cite{PappF}, de azt jóval túlszárnyalva olyan grammatikai információk nyújtásával, amelyekre későbbi számítógépes alkalmazásokban lehetett támaszkodni. Tanítványai és ez utóbbiak munkatársai a számítógépes nyelvészetet a generatív nyelvészettel ötvözve vizsgálták a nyelvi rekurzió megjelenését a kognícióban \cite{Hunyadi}.

A MTA-SZTE Mesterséges Intelligencia kutatócsoportban a gépi tanulási kutatások mellett jelentős eredmények születtek az MI alklmazások területein is.  
Tóth László a kutatócsoport munkatársa a beszédfelismerés területén dolgozott. Kocsor Andrással kifejlesztették  a BeszédMester beszédterápiás és olvasást segítő programot kisiskolások számára \cite{Kocsor}. Neuronhálós módszereket fejlesztettek beszédfelismerésre  szegmens-alapú hibrid modellekben \cite{Toth1}, és  tandem rendszerekben \cite{Toth2}.
Csirik János  szerzőtársával az alakfelismerés területén ért el jelentős eredményeket.  A futás hossz-kódolás egy  tömörítési eljárás, amellyel egy karaktersorozatban a hosszasan ismétlődő karaktereket egyetlen értékként és számként tárolják, az eredeti teljes karaktersorozat helyett. A  szerzők egy hatékony módszert dolgoztak ki  a futáshossz kóddal megadott karaktersorozatok távolságának meghatározására \cite{Bunke1}. A szerzők egy másik cikkükben a szerkesztési távolság olyan általánosításával foglalkoztak, amelyben a törlés és beszúrás költsége egy konstans, a csere költsége pedig egy paraméter.  Hatékony algoritmust adtak a szerkesztési távolság meghatározására \cite{Bunke2}. 
Gyimóthy Tibor munkatársaival gépi tanulási módszerekkel valódi szoftver rendszerekre igazolták, hogy egyszerűen számítható szoftver metrikák segítségével jól lehet becsülni a program kódok minőségét \cite{Gyimothy}. 
A kutatócsoport munkatársai létrehozták az 1.2 millió szavas annotált magyar nyelvi korpuszt (Szeged Korpusz), amelynek meghatározó szerepe volt a hazai számítógépes nyelvészeti kutatásokban \cite{Alexin2}.  Ugyancsak a kutatócsoport munkatársai indították el 2003-ban  Magyar Számítógépes Nyelvészeti Konferenciát, amely azóta minden évben a hazai nyelvészeti kutatási eredmények ismertetésének jelentős fóruma.

\section{Modern elméleti MI kutatások Magyarországon}
A mesterséges intelligencia (MI) egy nagyon általános és nehezen körülhatárolható fogalom, amely magában foglalja a gépi tanulást, ezen belül pedig a mélytanulást, amely manapság különösen nagy figyelmet kap. Tágabb nézőpontból azonban minden matematikai módszer, amelynek célja adathalmazokban és struktúrákban mintázatokat, összefüggéseket keresni, illetve hiányzó információkat jósolni, az MI részének tekinthető. Ez magyarázza, hogy az MI elméleti alapjai között fontos szerepet kap a statisztika, a valószínűségszámítás, és kevésbé ismert módon a statisztikus fizika is. Egy másik lényeges elméleti pillér a többváltozós optimalizáció, amely az MI-modellek tanításához kapcsolódik. A mélytanulásra épülő algoritmusok elterjedésével pedig egyre fontosabbá váltak azok a kutatások is, amelyek ezen modellek magyarázhatóságát, sebezhetőségét és általában véve működésüket vizsgálják. Mindezekről az irányokról ebben a fejezetben úgy fogunk írni, hogy különösen kiemeljük a magyar vonatkozásokat.

A 2010-es években számos (részben váratlan) áttörés történt a mélytanulás területén, amelyek az MI-forradalom első jeleiként tekinthetők. Ennek az időszaknak az elejére tehető, hogy a mélytanulás keretrendszerében számos lényeges elem letisztult, mint például az aktivációs függvények jelentősége, illetve a regularizációs és optimalizációs módszerek különböző típusai. Az MI-forradalom számos nagy áttörésének magyar vonatkozása is van. Ezek részben magyar származású kutatók által külföldön elért eredmények, illetve hazai kutatások eredményei.

2013-ban Szegedy Krisztián és társszerzői megmutatták, hogy a mély neurális hálózatok átverhetők \cite{szegedy2014}. Példájukban vizuálisan észrevehetetlen módon tudtak perturbálni képeket úgy, hogy egy adott modell azokat látványosan rosszul ismerje fel. Ez a cikk egy teljes kutatási területet alapozott meg, amely a neurális hálózatok sebezhetőségével foglalkozik, és ezért elnyerte az ,,ICLR 2024 Test of Time'' díjat is. A témakör kutatása hazai intézményekben is folytatódott. A Szegedi Tudományegyetemen Jelasity Márk és társszerzői több meglepő eredményt publikáltak. Az egyik ilyen cikkben \cite{Jel1} megmutatták, hogy több különböző modell egyszerre is átverhető ugyanazzal a perturbációval egy adott képlista esetén, ráadásul a tévedés típusát is irányítani tudták. Későbbi cikkeikben \cite{Jel2} pedig a különböző támadások elleni robusztusság témáját is körüljárták. A Pázmány Péter Katolikus Egyetemen Horváth András és Csaba M. Józsa a Neural Radiance Fields (NeRFs) módszer elleni lehetséges támadásokat vizsgálták \cite{Horvath_2023_ICCV}. Ez a módszer kétdimenziós képekből képes háromdimenziós reprezentációt létrehozni, amelyet a robotikában és a kiterjesztett valóságban is alkalmaznak.

A Google DeepMind által kifejlesztett AlphaGo sikertörténetének is több magyar vonatkozása van. Ez a program 2015-ben elsőként győzött le profi gójátékost, és ezzel a váratlan eredménnyel a témában laikusok figyelmét is az MI-technológiára irányította. Az AlphaGo hatékonysága két fontos pilléren nyugszik. Az első a Monte-Carlo fa keresési (Monte-Carlo Tree Search, MCTS) módszer finomítása, amely a már korábban említett Kocsis Levente (SZTAKI) és Szepesvári Csaba által jegyzett híres ,,Bandit Based Monte-Carlo Planning'' cikkre épül. A másik alappillér a mély vizuális, konvolúciós hálózatok tanításának fejlődése, amelyben fontos szerepet játszik a Szegedy Krisztián által kidolgozott batch normalization módszer \cite{batchnorm}. Az AlphaGo sikere látványosan illusztrálja a megerősítéses tanulási módszerek (Reinforcement Learning, RL) hatékonyságát. Nem véletlen, hogy ez a témakör is hatalmas fejlődésen ment keresztül viszonylag rövid idő alatt. Magyar kutatók közül még kiemelkednek Neu Gergely eredményei az RL terén \cite{Neu2007ApprenticeshipLU}, amelyek a téma elméleti alapjaihoz járulnak hozzá.

A 2010-ben elindult MI-forradalom számos izgalmas fordulatot és paradigmaváltást hozott. Az egyik ilyen meglepő eredmény, hogy a túlparametrizált hálózatok mégsem hajlamosak a túltanulásra, ahogy azt korábban feltételezték. Ez az észrevétel rengeteg új kutatást inspirált. Egy sokat hivatkozott cikkben \cite{du2019} Póczos Barnabás és társszerzői bebizonyították, hogy az ilyen hálózatokon a ,,gradient descent'' algoritmus az optimumhoz konvergál. Szintén részben Póczos Barnabás nevéhez köthető az MMD-GAN modellek megalkotása \cite{NIPS2017MMDGAN}, amelyek ötvözik a GMMN-ek (Generative Moment Matching Network) és a GAN-ok (Generative Adversarial Network) előnyeit.

A túlparametrizált hálózatok hatékonysága is motiválja azt az elméleti kutatási irányt is, amelyben növekvő neurális hálózatok határértékét vizsgálják annak reményében, hogy a limeszben letisztultabb modellt kaphassunk. Ennek a gondolatnak több megvalósulása is van. Az egyik ilyen, amely a gráf neurális hálózatok (GNN) limeszét vizsgálja, közvetlenül épít a Christian Borges, Jennifer Chayes, Lovász Laszlo, T Sós Vera, Szegedy Balazs, Vesztergombi Katalin által \cite{Glim_2006} elindított sűrű gráfok limeszelméletére, illetve a Backhausz Ágnes és Szegedy Balázs által kidolgozott ,,action convergence'' általános limeszelméletre.

A mesterséges intelligencia matematikai alapjainak tisztázásához minden olyan állítás hozzájárul, amely a mélytanulás valamilyen fajta robusztusságát igazolja. A Rényi intézetben Csiszárik Adrián, Kőrösi-Szabó Péter, Matszangosz Ákos, Papp Gergely és Varga Dániel olyan módszertant dolgozott ki \cite{NEURIPS2021matching}, amellyel különböző betanított modellek belső reprezentációinak az inicializálástól való függetlenségét tudták vizsgálni, és jelentős robusztussági tulajdonságokat fedeztek fel.

A neurális hálózatok döntéshozatali folyamata gyakran nehezen érthető és követhető, még a fejlesztők számára is. Ezt hívják fekete doboz tulajdonságnak. Az interpretálhatóság kérdése az, hogy milyen mértékben lehet értelmezni és megmagyarázni a hálózat által hozott döntéseket vagy előrejelzéseket. Ebben a témában ért el fontos eredményeket az SZTE kutatója, Turán György, aki többféle kontextusban is vizsgálta az interpretálhatóság kérdését. Az egyik ilyen ígéretes irány a mélytanulás segítségével generált hibajavító kódok interpretálhatósága \cite{errcorr}.

A gépi tanulás matematikai alapjainak kutatásában az egyik fő terület a statisztikai tanuláselmélet. Ide tartozik a gépi tanulásban és rendszerazonosításban fellépő sztochasztikus modellek valószínűségelméleti és statisztikai vizsgálata. Ennek a témának a hazai szakértője Csáji Balázs (SZTAKI), aki a kernel módszerek és a konfidencia tartományok vizsgálatában is kiemelkedő eredményeket ért el \cite{Csaji1},\cite{Csaji2}.

Fontosnak tartjuk megemlíteni, hogy van több jelenleg előkészületben lévő, illetve nemrég kezdődött nagyobb interdiszciplináris kutatási projekt is, amelyek a mesterséges intelligencia új megközelítéseit, illetve felhasználásait keresik. Az egyik ilyen projekt a DYNASNET ERC szinergia program keretében zajlik a Rényi intézetben, amelynek koordinátorai Barabási Albert László, Lovász László és Jaroslav Nešetřil. Ennek a kutatásnak a célja, hogy a hálózatelmélet eredményeit közelebb hozza a neurális hálózatok világához \cite{posfai2024impact}. Egy másik, Barabási Albert László nevéhez köthető irány a táplálkozástudomány szilárdabb alapokra helyezését célozza meg gépi tanulás segítségével.

\section{Modern MI Alkalmazások}

A MI már kiforrott, klasszikus területének számítanak az ajánló rendszerek, amely területen hazánkat a Tikk Domokos vezette csapat tette világhíressé, amely a Netflix versenyén hajszállal maradt le az egymillió dolláros fődíjról \cite{takacs2009scalable}. Hozzájuk fűződik a rekurrens neuronháló alapú ajánlórendszerek \cite{hidasi2015session} vizsgálata, amellyel új kutatási területet nyitottak.

Hasonlóképpen klasszikus, és hazánkban nagyon erős terület a hálózatok elemzése \cite{posfai2024impact}. A hálózatkutatás legújabb trendje a vektortér beágyazások alkalmazása \cite{rozemberczki2021pytorch}, amelyeket alkalmazunk olyan teljesen eltérő területeken, mint a közösségi médiában oltásokkal kapcsolatos vélemények terjedése \cite{beres2023network} vagy a kriptopénzek anonimitás-vizsgálata \cite{beres2021blockchain}. 

A nyelvtechnológia területén fontos eredményeket értek el a Nyelvtudományi Intézet munkatársai \cite{ligeti2024hulu}, valamint Kornai András csoportja \cite{acs2023morphosyntactic}. A Társadalomtudományi Kutatóközpontban fontos kapcsolódó kutatási terület a politikai szövegek elemzése \cite{sebok2021multiclass}.

A gépi látás területén belül erős a hazai geometriai modellezéssel \cite{barath2021graph}, pontfelhő alapú követéssel \cite{nagy2021changegan} kapcsolatos iskola és az alkalmazott kutatások \cite{szemenyei2022fully}. Az autonóm közlekedés alkalmazásaiban a  gépi látás \cite{zsedrovits2016onboard} mellett a megerősítés tanulás \cite{antal2023backflipping} is kiemelt szerepet kap.

Az egészségügy területén mind a képalkotó diagnosztikában \cite{ribli2018detecting}, mind az elektronikus egészség-adatok, laboreredmények vizsgálatában \cite{juhasz2024blood} több hazai orvosi egyetem \cite{paragh2018identifying} és klinika partner. Új irány az öregedés-kutatás genetikusok (Vellai Tibor), orvosok és bioinformatikusok (Kerepesi Csaba) vezetésével \cite{kerepesi2018prediction}. Megerősítéses tanulást használva sikerült automatizált őssejt gyártó platform irányítási rendszerét optimalizálni \cite{egri2020bio}, valamint a gyógyszergyártás támogatására gépi látást alkalmazni \cite{nagy2023interpretable}. Fontos kutatási terület a fordított irány, a biológiai neuronhálózatok viselkedésének vizsgálata, amellyel nemzetközi hírű agykutatóink foglalkoznak \cite{hegedus2023cholinergic}.

Gyártás, robotika, ipar 4.0 területén kiemelkedő a Monostori László nevével fémjelzett iskola \cite{gao2024artificial}. Kutatásaink legfontosabb partnerei a hazai gyártó nagyvállalatok: az Audi, Bosch \cite{mandli2017time}, Opel \cite{viharos2021adaptive} és a Knorr-Bremse. A BME-n Gépi látás \cite{geier2022method, shaheen2023data} és a mért adatok modellezése \cite{fothi2024deep} segítségével precíziós gyártási feladatokra született megoldás \cite{adizue2023surface}. Az Eötvös Loránd Tudományegyetem Bosch által alapított tanszékén többek között ember-gép együttműködéssel kapcsolatos kutatásokat folytat Lőrincz András csoportja \cite{fothi2024deep}. 

Az MI távközlési alkalmazásaival kapcsolatos kutatásokat a budapesti Ericsson \cite{kelen2023theoretical} és Nokia \cite{horvath2023sub} fejlesztő központok vezetésével végezzük. A következő, 5G utáni mobil hálózatok architektúrájának központi komponense lesz a MI megoldások támogatása, amelynek kutatásába bekapcsolódunk \cite{merluzzi2023hexa}.

Az adatok elemzése, modellezése, előrejelzése, magyarázatok, okok keresése megjelenik az enerigahálózatok, megújuló energia termelés \cite{csaji2020sampling}, épület \cite{kurent2024bayesian} és földszerkezetek \cite{li2022sensitivity}, az agrárium \cite{nyeki2021application}, és az oktatás \cite{berezvai2021can} területein is.

\bibliography{template}

\begin{thebibliography}{116}
\expandafter\ifx\csname natexlab\endcsname\relax\def\natexlab#1{#1}\fi
\providecommand{\url}[1]{\texttt{#1}}
\providecommand{\href}[2]{#2}
\providecommand{\path}[1]{#1}
\providecommand{\DOIprefix}{doi:}
\providecommand{\ArXivprefix}{arXiv:}
\providecommand{\URLprefix}{URL: }
\providecommand{\Pubmedprefix}{pmid:}
\providecommand{\doi}[1]{\href{http://dx.doi.org/#1}{\path{#1}}}
\providecommand{\Pubmed}[1]{\href{pmid:#1}{\path{#1}}}
\providecommand{\bibinfo}[2]{#2}
\ifx\xfnm\relax \def\xfnm[#1]{\unskip,\space#1}\fi
%Type = Article
\bibitem[{Acs et~al.(2023)Acs, Hamerlik, Schwartz, Smith \& Kornai}]{acs2023morphosyntactic}
\bibinfo{author}{Acs, J.}, \bibinfo{author}{Hamerlik, E.}, \bibinfo{author}{Schwartz, R.}, \bibinfo{author}{Smith, N.~A.}, \& \bibinfo{author}{Kornai, A.} (\bibinfo{year}{2023}).
\newblock \bibinfo{title}{Morphosyntactic probing of multilingual bert models}.
\newblock {\it \bibinfo{journal}{Natural Language Engineering}\/},  (pp. \bibinfo{pages}{1--40}).
%Type = Article
\bibitem[{Adizue et~al.(2023)Adizue, Tura, Isaya, Farkas \& Tak{\'a}cs}]{adizue2023surface}
\bibinfo{author}{Adizue, U.~L.}, \bibinfo{author}{Tura, A.~D.}, \bibinfo{author}{Isaya, E.~O.}, \bibinfo{author}{Farkas, B.~Z.}, \& \bibinfo{author}{Tak{\'a}cs, M.} (\bibinfo{year}{2023}).
\newblock \bibinfo{title}{Surface quality prediction by machine learning methods and process parameter optimization in ultra-precision machining of aisi d2 using cbn tool}.
\newblock {\it \bibinfo{journal}{The International Journal of Advanced Manufacturing Technology}\/},  {\it \bibinfo{volume}{129}\/}, \bibinfo{pages}{1375--1394}.
%Type = Inproceedings
\bibitem[{Alexin et~al.(2003)Alexin, Csirik, Gyimóthy, Bibok, Hatvani, Prószéky \& Tihanyi}]{Alexin1}
\bibinfo{author}{Alexin, Z.}, \bibinfo{author}{Csirik, J.}, \bibinfo{author}{Gyimóthy, T.}, \bibinfo{author}{Bibok, K.}, \bibinfo{author}{Hatvani, C.}, \bibinfo{author}{Prószéky, G.}, \& \bibinfo{author}{Tihanyi, L.} (\bibinfo{year}{2003}).
\newblock \bibinfo{title}{Manually annotated hungarian corpus}.
\newblock In \bibinfo{editor}{A.~Copestake}, \& \bibinfo{editor}{J.~Hajic} (Eds.), {\it \bibinfo{booktitle}{Proceedings of the 10th Conference of the European Chapter of the Association for Computational Linguistics (EACL)}\/} (pp. \bibinfo{pages}{53--56}).
\newblock \bibinfo{address}{Stroudsburg, PA, USA}: \bibinfo{publisher}{Association for Computational Linguistics (ACL)}.
%Type = Inproceedings
\bibitem[{Alexin et~al.(1996)Alexin, Gyimóthy \& Boström}]{Alexin2}
\bibinfo{author}{Alexin, Z.}, \bibinfo{author}{Gyimóthy, T.}, \& \bibinfo{author}{Boström, H.} (\bibinfo{year}{1996}).
\newblock \bibinfo{title}{Integrating algorithmic debugging and unfolding transformations in an interactive learner}.
\newblock In \bibinfo{editor}{W.~Wahlster} (Ed.), {\it \bibinfo{booktitle}{ECAI 96: Proceedings of the 12th European Conference on Artificial Intelligence}\/} (pp. \bibinfo{pages}{403--408}).
\newblock \bibinfo{address}{Chichester, United Kingdom}: \bibinfo{publisher}{John Wiley \& Sons} volume \bibinfo{volume}{716}.
%Type = Inproceedings
\bibitem[{Andréka et~al.(1975)Andréka, Gergely \& Németi}]{Andréka}
\bibinfo{author}{Andréka, H.}, \bibinfo{author}{Gergely, T.}, \& \bibinfo{author}{Németi, I.} (\bibinfo{year}{1975}).
\newblock \bibinfo{title}{Definition theory as basis for a creative problem solver}.
\newblock In {\it \bibinfo{booktitle}{Advance Papers of the Fourth International Joint Conference on Artificial Intelligence}\/} (pp. \bibinfo{pages}{40--45}).
\newblock \bibinfo{address}{Massachusetts}: \bibinfo{publisher}{MIT Publications Department}.
%Type = Article
\bibitem[{Angluin et~al.(1997)Angluin, Krikis, Sloan \& Turán}]{Angluin}
\bibinfo{author}{Angluin, D.}, \bibinfo{author}{Krikis, M.}, \bibinfo{author}{Sloan, R.~H.}, \& \bibinfo{author}{Turán, G.} (\bibinfo{year}{1997}).
\newblock \bibinfo{title}{Malicious omissions and errors in answers to membership queries}.
\newblock {\it \bibinfo{journal}{Machine Learning}\/},  {\it \bibinfo{volume}{28}\/}, \bibinfo{pages}{211--255}.
%Type = Article
\bibitem[{Antal et~al.(2023)Antal, P{\'e}ni \& T{\'o}th}]{antal2023backflipping}
\bibinfo{author}{Antal, P.}, \bibinfo{author}{P{\'e}ni, T.}, \& \bibinfo{author}{T{\'o}th, R.} (\bibinfo{year}{2023}).
\newblock \bibinfo{title}{Backflipping with miniature quadcopters by gaussian-process-based control and planning}.
\newblock {\it \bibinfo{journal}{IEEE Transactions on Control Systems Technology}\/}, .
%Type = Article
\bibitem[{Barath \& Matas(2021)}]{barath2021graph}
\bibinfo{author}{Barath, D.}, \& \bibinfo{author}{Matas, J.} (\bibinfo{year}{2021}).
\newblock \bibinfo{title}{Graph-cut ransac: Local optimization on spatially coherent structures}.
\newblock {\it \bibinfo{journal}{IEEE transactions on pattern analysis and machine intelligence}\/},  {\it \bibinfo{volume}{44}\/}, \bibinfo{pages}{4961--4974}.
%Type = Article
\bibitem[{B{\'e}res et~al.(2023)B{\'e}res, Michaletzky, Csoma \& Bencz{\'u}r}]{beres2023network}
\bibinfo{author}{B{\'e}res, F.}, \bibinfo{author}{Michaletzky, T.~V.}, \bibinfo{author}{Csoma, R.}, \& \bibinfo{author}{Bencz{\'u}r, A.~A.} (\bibinfo{year}{2023}).
\newblock \bibinfo{title}{Network embedding aided vaccine skepticism detection}.
\newblock {\it \bibinfo{journal}{Applied Network Science}\/},  {\it \bibinfo{volume}{8}\/}, \bibinfo{pages}{11}.
%Type = Inproceedings
\bibitem[{B{\'e}res et~al.(2021)B{\'e}res, Seres, Bencz{\'u}r \& Quintyne-Collins}]{beres2021blockchain}
\bibinfo{author}{B{\'e}res, F.}, \bibinfo{author}{Seres, I.~A.}, \bibinfo{author}{Bencz{\'u}r, A.~A.}, \& \bibinfo{author}{Quintyne-Collins, M.} (\bibinfo{year}{2021}).
\newblock \bibinfo{title}{Blockchain is watching you: Profiling and deanonymizing ethereum users}.
\newblock In {\it \bibinfo{booktitle}{2021 IEEE international conference on decentralized applications and infrastructures (DAPPS)}\/} (pp. \bibinfo{pages}{69--78}).
\newblock \bibinfo{organization}{IEEE}.
%Type = Article
\bibitem[{Berezvai et~al.(2021)Berezvai, Luk{\'a}ts \& Molontay}]{berezvai2021can}
\bibinfo{author}{Berezvai, Z.}, \bibinfo{author}{Luk{\'a}ts, G.~D.}, \& \bibinfo{author}{Molontay, R.} (\bibinfo{year}{2021}).
\newblock \bibinfo{title}{Can professors buy better evaluation with lenient grading? the effect of grade inflation on student evaluation of teaching}.
\newblock {\it \bibinfo{journal}{Assessment \& Evaluation in Higher Education}\/},  {\it \bibinfo{volume}{46}\/}, \bibinfo{pages}{793--808}.
%Type = Inproceedings
\bibitem[{Borgs et~al.(2006)Borgs, Chayes, Lovász, Sós, Szegedy \& Vesztergombi}]{Glim_2006}
\bibinfo{author}{Borgs, C.}, \bibinfo{author}{Chayes, J.}, \bibinfo{author}{Lovász, L.}, \bibinfo{author}{Sós, V.~T.}, \bibinfo{author}{Szegedy, B.}, \& \bibinfo{author}{Vesztergombi, K.} (\bibinfo{year}{2006}).
\newblock \bibinfo{title}{Graph limits and parameter testing}.
\newblock In {\it \bibinfo{booktitle}{Proceedings of the thirty-eighth annual ACM symposium on Theory of Computing}\/} (p. \bibinfo{pages}{261–270}).
\newblock \bibinfo{publisher}{ACM} volume~\bibinfo{volume}{75} of {\it \bibinfo{series}{STOC06}\/}.
\newblock \URLprefix \url{http://dx.doi.org/10.1145/1132516.1132556}. \DOIprefix\doi{10.1145/1132516.1132556}.
%Type = Article
\bibitem[{Bunke \& Csirik(1995{\natexlab{a}})}]{Bunke1}
\bibinfo{author}{Bunke, H.}, \& \bibinfo{author}{Csirik, J.} (\bibinfo{year}{1995}{\natexlab{a}}).
\newblock \bibinfo{title}{An improved algorithm for computing the edit distance of run-length coded strings}.
\newblock {\it \bibinfo{journal}{Information Processing Letters}\/},  {\it \bibinfo{volume}{54}\/}, \bibinfo{pages}{93--96}.
%Type = Article
\bibitem[{Bunke \& Csirik(1995{\natexlab{b}})}]{Bunke2}
\bibinfo{author}{Bunke, H.}, \& \bibinfo{author}{Csirik, J.} (\bibinfo{year}{1995}{\natexlab{b}}).
\newblock \bibinfo{title}{Parametric string edit distance and its application to pattern recognition}.
\newblock {\it \bibinfo{journal}{IEEE Transactions on Systems, Man, and Cybernetics}\/},  {\it \bibinfo{volume}{25}\/}, \bibinfo{pages}{202--206}.
%Type = Inproceedings
\bibitem[{Calderwood \& Szeredi(1989)}]{Calderwood}
\bibinfo{author}{Calderwood, A.}, \& \bibinfo{author}{Szeredi, P.} (\bibinfo{year}{1989}).
\newblock \bibinfo{title}{Scheduling or-parallelism in aurora - the manchester scheduler}.
\newblock In {\it \bibinfo{booktitle}{Proceedings of the Sixth International Conference on Logic Programming}\/} (pp. \bibinfo{pages}{419--435}).
\newblock \bibinfo{publisher}{The MIT Press}.
%Type = Article
\bibitem[{Chrobak et~al.(2000)Chrobak, Lőrincz \& Buzsáki}]{Chrobak}
\bibinfo{author}{Chrobak, J.~J.}, \bibinfo{author}{Lőrincz, A.}, \& \bibinfo{author}{Buzsáki, G.} (\bibinfo{year}{2000}).
\newblock \bibinfo{title}{Physiological patterns in the hippocampo‐entorhinal cortex system}.
\newblock {\it \bibinfo{journal}{Hippocampus}\/},  {\it \bibinfo{volume}{10}\/}, \bibinfo{pages}{457--465}.
%Type = Article
\bibitem[{Chua \& Roska(1993)}]{Chua}
\bibinfo{author}{Chua, L.~O.}, \& \bibinfo{author}{Roska, T.} (\bibinfo{year}{1993}).
\newblock \bibinfo{title}{The cnn paradigm}.
\newblock {\it \bibinfo{journal}{IEEE Transactions on Circuits and Systems I - Fundamental Theory and Applications}\/},  {\it \bibinfo{volume}{40}\/}, \bibinfo{pages}{147--156}.
%Type = Article
\bibitem[{Cs{\'a}ji \& Kis(2019)}]{Csaji2}
\bibinfo{author}{Cs{\'a}ji, B.~C.}, \& \bibinfo{author}{Kis, K.~B.} (\bibinfo{year}{2019}).
\newblock \bibinfo{title}{Distribution-free uncertainty quantification for kernel methods by gradient perturbations}.
\newblock {\it \bibinfo{journal}{MACHINE LEARNING}\/},  {\it \bibinfo{volume}{108}\/}, \bibinfo{pages}{1677--1699}. \URLprefix \url{https://eprints.sztaki.hu/9712/}.
%Type = Article
\bibitem[{Cs{\'a}ji et~al.(2020)Cs{\'a}ji, Kis \& Kov{\'a}cs}]{csaji2020sampling}
\bibinfo{author}{Cs{\'a}ji, B.~C.}, \bibinfo{author}{Kis, K.~B.}, \& \bibinfo{author}{Kov{\'a}cs, A.} (\bibinfo{year}{2020}).
\newblock \bibinfo{title}{A sampling-and-discarding approach to stochastic model predictive control for renewable energy systems}.
\newblock {\it \bibinfo{journal}{IFAC-PapersOnLine}\/},  {\it \bibinfo{volume}{53}\/}, \bibinfo{pages}{7142--7147}.
%Type = Inproceedings
\bibitem[{Csisz\'{a}rik et~al.(2021)Csisz\'{a}rik, K\H{o}r\"{o}si-Szab\'{o}, Matszangosz, Papp \& Varga}]{NEURIPS2021matching}
\bibinfo{author}{Csisz\'{a}rik, A.}, \bibinfo{author}{K\H{o}r\"{o}si-Szab\'{o}, P.}, \bibinfo{author}{Matszangosz, A.}, \bibinfo{author}{Papp, G.}, \& \bibinfo{author}{Varga, D.} (\bibinfo{year}{2021}).
\newblock \bibinfo{title}{Similarity and matching of neural network representations}.
\newblock In \bibinfo{editor}{M.~Ranzato}, \bibinfo{editor}{A.~Beygelzimer}, \bibinfo{editor}{Y.~Dauphin}, \bibinfo{editor}{P.~Liang}, \& \bibinfo{editor}{J.~W. Vaughan} (Eds.), {\it \bibinfo{booktitle}{Advances in Neural Information Processing Systems}\/} (pp. \bibinfo{pages}{5656--5668}).
\newblock \bibinfo{publisher}{Curran Associates, Inc.} volume~\bibinfo{volume}{34}.
\newblock \URLprefix \url{https://proceedings.neurips.cc/paper_files/paper/2021/file/2cb274e6ce940f47beb8011d8ecb1462-Paper.pdf}.
%Type = Article
\bibitem[{Csuhaj-Varjú \& Dassow(1990)}]{CsuhajVarju2}
\bibinfo{author}{Csuhaj-Varjú, E.}, \& \bibinfo{author}{Dassow, J.} (\bibinfo{year}{1990}).
\newblock \bibinfo{title}{On cooperating/distributed grammar systems}.
\newblock {\it \bibinfo{journal}{Journal of Information Processing and Cybernetics}\/},  {\it \bibinfo{volume}{26}\/}, \bibinfo{pages}{49--63}.
%Type = Book
\bibitem[{Csuhaj-Varjú et~al.(1994)Csuhaj-Varjú, Dassow, Kelemen \& Păun}]{CsuhajVarju1}
\bibinfo{author}{Csuhaj-Varjú, E.}, \bibinfo{author}{Dassow, J.}, \bibinfo{author}{Kelemen, J.}, \& \bibinfo{author}{Păun, G.} (\bibinfo{year}{1994}).
\newblock {\it \bibinfo{title}{Grammar systems: A grammatical approach to distribution and cooperation}\/}.
\newblock \bibinfo{address}{Yverdon, Switzerland}: \bibinfo{publisher}{Gordon and Breach Science Publishers}.
%Type = Article
\bibitem[{Csáji et~al.(2015)Csáji, Campi \& Weyer}]{Csaji1}
\bibinfo{author}{Csáji, B.~C.}, \bibinfo{author}{Campi, M.~C.}, \& \bibinfo{author}{Weyer, E.} (\bibinfo{year}{2015}).
\newblock \bibinfo{title}{Sign-perturbed sums: A new system identification approach for constructing exact non-asymptotic confidence regions in linear regression models}.
\newblock {\it \bibinfo{journal}{IEEE Transactions on Signal Processing}\/},  {\it \bibinfo{volume}{63}\/}, \bibinfo{pages}{169--181}. \DOIprefix\doi{10.1109/TSP.2014.2369000}.
%Type = Article
\bibitem[{Csáji et~al.(2006)Csáji, Monostori \& Kádár}]{Csaji}
\bibinfo{author}{Csáji, B.~C.}, \bibinfo{author}{Monostori, L.}, \& \bibinfo{author}{Kádár, B.} (\bibinfo{year}{2006}).
\newblock \bibinfo{title}{Reinforcement learning in a distributed market-based production control system}.
\newblock {\it \bibinfo{journal}{Advanced Engineering Informatics}\/},  {\it \bibinfo{volume}{20}\/}, \bibinfo{pages}{279--288}. \DOIprefix\doi{10.1016/j.aei.2006.01.001}.
%Type = Inproceedings
\bibitem[{Devroye et~al.(2022)Devroye, Mohammadi, Mulgund, Naik, Shekhar, Turán, Wei \& Žefran}]{errcorr}
\bibinfo{author}{Devroye, N.}, \bibinfo{author}{Mohammadi, N.}, \bibinfo{author}{Mulgund, A.}, \bibinfo{author}{Naik, H.}, \bibinfo{author}{Shekhar, R.}, \bibinfo{author}{Turán, G.}, \bibinfo{author}{Wei, Y.}, \& \bibinfo{author}{Žefran, M.} (\bibinfo{year}{2022}).
\newblock \bibinfo{title}{Interpreting deep-learned error-correcting codes}.
\newblock In {\it \bibinfo{booktitle}{2022 IEEE International Symposium on Information Theory (ISIT)}\/} (pp. \bibinfo{pages}{2457--2462}).
\newblock \DOIprefix\doi{10.1109/ISIT50566.2022.9834599}.
%Type = Article
\bibitem[{Dombi(1982)}]{Dombi1}
\bibinfo{author}{Dombi, J.} (\bibinfo{year}{1982}).
\newblock \bibinfo{title}{A general-class of fuzzy operators, the demorgan class of fuzzy operators and fuzziness measures induced by fuzzy operators}.
\newblock {\it \bibinfo{journal}{Fuzzy Sets and Systems}\/},  {\it \bibinfo{volume}{8}\/}, \bibinfo{pages}{149--163}.
%Type = Article
\bibitem[{Dombi(2008)}]{Dombi2}
\bibinfo{author}{Dombi, J.} (\bibinfo{year}{2008}).
\newblock \bibinfo{title}{Towards a general class of operators for fuzzy systems}.
\newblock {\it \bibinfo{journal}{IEEE Transactions on Fuzzy Systems}\/},  {\it \bibinfo{volume}{16}\/}, \bibinfo{pages}{477--484}.
%Type = Misc
\bibitem[{Du et~al.(2019)Du, Zhai, Poczos \& Singh}]{du2019}
\bibinfo{author}{Du, S.~S.}, \bibinfo{author}{Zhai, X.}, \bibinfo{author}{Poczos, B.}, \& \bibinfo{author}{Singh, A.} (\bibinfo{year}{2019}).
\newblock \bibinfo{title}{Gradient descent provably optimizes over-parameterized neural networks}.
\newblock \URLprefix \url{https://arxiv.org/abs/1810.02054}. \href{http://arxiv.org/abs/1810.02054}{\tt arXiv:1810.02054}.
%Type = Misc
\bibitem[{Dömölki(2023)}]{Domolki}
\bibinfo{author}{Dömölki, B.} (\bibinfo{year}{2023}).
\newblock \bibinfo{title}{Mesterséges intelligencia múltja {Magyarországon}}.
\newblock \bibinfo{howpublished}{\url{https://itf.njszt.hu/wp-content/uploads/2023/04/MIMM-5.pdf}}.
%Type = Article
\bibitem[{Egri et~al.(2020)Egri, Cs{\'a}ji, Kis, Monostori, V{\'a}ncza, Ochs, Jung, K{\"o}nig, Schmitt, Brecher et~al.}]{egri2020bio}
\bibinfo{author}{Egri, P.}, \bibinfo{author}{Cs{\'a}ji, B.~C.}, \bibinfo{author}{Kis, K.~B.}, \bibinfo{author}{Monostori, L.}, \bibinfo{author}{V{\'a}ncza, J.}, \bibinfo{author}{Ochs, J.}, \bibinfo{author}{Jung, S.}, \bibinfo{author}{K{\"o}nig, N.}, \bibinfo{author}{Schmitt, R.}, \bibinfo{author}{Brecher, C.} et~al. (\bibinfo{year}{2020}).
\newblock \bibinfo{title}{Bio-inspired control of automated stem cell production}.
\newblock {\it \bibinfo{journal}{Procedia CIRP}\/},  {\it \bibinfo{volume}{88}\/}, \bibinfo{pages}{600--605}.
%Type = Article
\bibitem[{Ekler et~al.(2009)Ekler, Kelényi, Dévai, Bakos \& Kiss}]{Ekler}
\bibinfo{author}{Ekler, P.}, \bibinfo{author}{Kelényi, I.}, \bibinfo{author}{Dévai, I.}, \bibinfo{author}{Bakos, B.}, \& \bibinfo{author}{Kiss, A.} (\bibinfo{year}{2009}).
\newblock \bibinfo{title}{Hybrid peer-to-peer content sharing in mobile networks}.
\newblock {\it \bibinfo{journal}{Journal of Networks}\/},  {\it \bibinfo{volume}{4}\/}, \bibinfo{pages}{119--132}.
%Type = Inproceedings
\bibitem[{Forstner et~al.(2005)Forstner, Lengyel, Levendovszky, Kelényi \& Charaf}]{Forstner2}
\bibinfo{author}{Forstner, B.}, \bibinfo{author}{Lengyel, L.}, \bibinfo{author}{Levendovszky, T.}, \bibinfo{author}{Kelényi, I.}, \& \bibinfo{author}{Charaf, H.} (\bibinfo{year}{2005}).
\newblock \bibinfo{title}{Supporting rapid application development on symbian platform}.
\newblock In {\it \bibinfo{booktitle}{Eurocon 2005: The International Conference on Computer as a Tool, Vol 1 and 2}\/} (pp. \bibinfo{pages}{72--75}).
\newblock \bibinfo{address}{New York, USA}: \bibinfo{publisher}{IEEE}.
%Type = Inproceedings
\bibitem[{Forstner et~al.(2006)Forstner, Lengyel, Levendovszky, Mezei, Kelényi \& Charaf}]{Forstner1}
\bibinfo{author}{Forstner, B.}, \bibinfo{author}{Lengyel, L.}, \bibinfo{author}{Levendovszky, T.}, \bibinfo{author}{Mezei, G.}, \bibinfo{author}{Kelényi, I.}, \& \bibinfo{author}{Charaf, H.} (\bibinfo{year}{2006}).
\newblock \bibinfo{title}{Model-based system development for embedded mobile platforms}.
\newblock In \bibinfo{editor}{R.~Machado}, \bibinfo{editor}{J.~Fernandes}, \bibinfo{editor}{M.~Riebisch}, \& \bibinfo{editor}{B.~Schatz} (Eds.), {\it \bibinfo{booktitle}{Joint Meeting of the Fourth Workshop on Model-Based Development of Computer-Based Systems and Third International Workshop on Model-Based Methodologies for Pervasive and Embedded Software, Proceedings}\/} (pp. \bibinfo{pages}{43--52}).
\newblock \bibinfo{address}{Los Alamitos, CA, USA}: \bibinfo{publisher}{IEEE Computer Society}.
%Type = Article
\bibitem[{F{\'o}thi et~al.(2024)F{\'o}thi, Skaf, Lu \& Fenech}]{fothi2024deep}
\bibinfo{author}{F{\'o}thi, {\'A}.}, \bibinfo{author}{Skaf, J.}, \bibinfo{author}{Lu, F.}, \& \bibinfo{author}{Fenech, K.} (\bibinfo{year}{2024}).
\newblock \bibinfo{title}{Deep nrsfm for multi-view multi-body pose estimation}.
\newblock {\it \bibinfo{journal}{Pattern Recognition Letters}\/}, .
%Type = Article
\bibitem[{Gao et~al.(2024)Gao, Kr{\"u}ger, Merklein, M{\"o}hring \& V{\'a}ncza}]{gao2024artificial}
\bibinfo{author}{Gao, R.~X.}, \bibinfo{author}{Kr{\"u}ger, J.}, \bibinfo{author}{Merklein, M.}, \bibinfo{author}{M{\"o}hring, H.-C.}, \& \bibinfo{author}{V{\'a}ncza, J.} (\bibinfo{year}{2024}).
\newblock \bibinfo{title}{Artificial intelligence in manufacturing: State of the art, perspectives, and future directions}.
\newblock {\it \bibinfo{journal}{CIRP Annals}\/}, .
%Type = Article
\bibitem[{Gebel et~al.(1994)Gebel, Pethő \& Zimmer}]{Gebel}
\bibinfo{author}{Gebel, J.}, \bibinfo{author}{Pethő, A.}, \& \bibinfo{author}{Zimmer, H.} (\bibinfo{year}{1994}).
\newblock \bibinfo{title}{Computing integral points on elliptic curves}.
\newblock {\it \bibinfo{journal}{Acta Arithmetica}\/},  {\it \bibinfo{volume}{68}\/}, \bibinfo{pages}{171--192}.
%Type = Article
\bibitem[{Geier et~al.(2022)Geier, P{\'o}ka, Jacs{\'o} \& Pereszlai}]{geier2022method}
\bibinfo{author}{Geier, N.}, \bibinfo{author}{P{\'o}ka, G.}, \bibinfo{author}{Jacs{\'o}, {\'A}.}, \& \bibinfo{author}{Pereszlai, C.} (\bibinfo{year}{2022}).
\newblock \bibinfo{title}{A method to predict drilling-induced burr occurrence in chopped carbon fibre reinforced polymer (cfrp) composites based on digital image processing}.
\newblock {\it \bibinfo{journal}{Composites Part B: Engineering}\/},  {\it \bibinfo{volume}{242}\/}, \bibinfo{pages}{110054}.
%Type = Inproceedings
\bibitem[{Gergely(1971)}]{Gergely}
\bibinfo{author}{Gergely, T.} (\bibinfo{year}{1971}).
\newblock \bibinfo{title}{Probability models for computer systems}.
\newblock In {\it \bibinfo{booktitle}{Computers and Automata}\/} (pp. \bibinfo{pages}{251--264}).
\newblock \bibinfo{address}{Brooklyn, New York}: \bibinfo{publisher}{Polytechnic Press}.
%Type = Article
\bibitem[{Gyimóthy et~al.(2005)Gyimóthy, Ferenc \& Siket}]{Gyimothy}
\bibinfo{author}{Gyimóthy, T.}, \bibinfo{author}{Ferenc, R.}, \& \bibinfo{author}{Siket, I.} (\bibinfo{year}{2005}).
\newblock \bibinfo{title}{Empirical validation of object-oriented metrics on open source software for fault prediction}.
\newblock {\it \bibinfo{journal}{IEEE Transactions on Software Engineering}\/},  {\it \bibinfo{volume}{31}\/}, \bibinfo{pages}{897--910}.
%Type = Article
\bibitem[{Hajnal et~al.(1993)Hajnal, Maass, Pudlák, Szegedy \& Turán}]{Hajnal}
\bibinfo{author}{Hajnal, A.}, \bibinfo{author}{Maass, W.}, \bibinfo{author}{Pudlák, P.}, \bibinfo{author}{Szegedy, M.}, \& \bibinfo{author}{Turán, G.} (\bibinfo{year}{1993}).
\newblock \bibinfo{title}{Threshold circuits of bounded depth}.
\newblock {\it \bibinfo{journal}{Journal of Computer and System Sciences}\/},  {\it \bibinfo{volume}{46}\/}, \bibinfo{pages}{129--154}.
%Type = Misc
\bibitem[{Halmosi et~al.(2024)Halmosi, Mohos \& Jelasity}]{Jel2}
\bibinfo{author}{Halmosi, L.}, \bibinfo{author}{Mohos, B.}, \& \bibinfo{author}{Jelasity, M.} (\bibinfo{year}{2024}).
\newblock \bibinfo{title}{Evaluating the adversarial robustness of semantic segmentation: Trying harder pays off}.
\newblock \URLprefix \url{https://arxiv.org/abs/2407.09150}. \href{http://arxiv.org/abs/2407.09150}{\tt arXiv:2407.09150}.
%Type = Article
\bibitem[{Heged{\"u}s et~al.(2023)Heged{\"u}s, Sviatk{\'o}, Kir{\'a}ly, Mart{\'\i}nez-Bellver \& Hangya}]{hegedus2023cholinergic}
\bibinfo{author}{Heged{\"u}s, P.}, \bibinfo{author}{Sviatk{\'o}, K.}, \bibinfo{author}{Kir{\'a}ly, B.}, \bibinfo{author}{Mart{\'\i}nez-Bellver, S.}, \& \bibinfo{author}{Hangya, B.} (\bibinfo{year}{2023}).
\newblock \bibinfo{title}{Cholinergic activity reflects reward expectations and predicts behavioral responses}.
\newblock {\it \bibinfo{journal}{Iscience}\/},  {\it \bibinfo{volume}{26}\/}.
%Type = Article
\bibitem[{Helbing et~al.(2000)Helbing, Farkas \& Vicsek}]{Helbing}
\bibinfo{author}{Helbing, D.}, \bibinfo{author}{Farkas, I.}, \& \bibinfo{author}{Vicsek, T.} (\bibinfo{year}{2000}).
\newblock \bibinfo{title}{Simulating dynamical features of escape panic}.
\newblock {\it \bibinfo{journal}{Nature}\/},  {\it \bibinfo{volume}{407}\/}, \bibinfo{pages}{487--490}.
%Type = Article
\bibitem[{Hidasi et~al.(2015)Hidasi, Karatzoglou, Baltrunas \& Tikk}]{hidasi2015session}
\bibinfo{author}{Hidasi, B.}, \bibinfo{author}{Karatzoglou, A.}, \bibinfo{author}{Baltrunas, L.}, \& \bibinfo{author}{Tikk, D.} (\bibinfo{year}{2015}).
\newblock \bibinfo{title}{Session-based recommendations with recurrent neural networks}.
\newblock {\it \bibinfo{journal}{arXiv preprint arXiv:1511.06939}\/}, .
%Type = Inproceedings
\bibitem[{Horv\'ath \& J\'ozsa(2023)}]{Horvath_2023_ICCV}
\bibinfo{author}{Horv\'ath, A.}, \& \bibinfo{author}{J\'ozsa, C.~M.} (\bibinfo{year}{2023}).
\newblock \bibinfo{title}{Targeted adversarial attacks on generalizable neural radiance fields}.
\newblock In {\it \bibinfo{booktitle}{Proceedings of the IEEE/CVF International Conference on Computer Vision (ICCV) Workshops}\/} (pp. \bibinfo{pages}{3718--3727}).
%Type = Article
\bibitem[{Horv{\'a}th et~al.(2023)Horv{\'a}th, K{\'a}d{\'a}r \& Szil{\'a}gyi}]{horvath2023sub}
\bibinfo{author}{Horv{\'a}th, G.}, \bibinfo{author}{K{\'a}d{\'a}r, A.}, \& \bibinfo{author}{Szil{\'a}gyi, P.} (\bibinfo{year}{2023}).
\newblock \bibinfo{title}{The sub-sequence summary method for detecting anomalies in logs}.
\newblock {\it \bibinfo{journal}{IEEE Access}\/},  {\it \bibinfo{volume}{11}\/}, \bibinfo{pages}{37412--37423}.
%Type = Inproceedings
\bibitem[{Horvath et~al.(2004)Horvath, Valyon, Strausz, Pataki, Sragner, Lasztovicza \& Szekely}]{Horvath}
\bibinfo{author}{Horvath, G.}, \bibinfo{author}{Valyon, J.}, \bibinfo{author}{Strausz, G.}, \bibinfo{author}{Pataki, M.}, \bibinfo{author}{Sragner, L.}, \bibinfo{author}{Lasztovicza, L.}, \& \bibinfo{author}{Szekely, N.} (\bibinfo{year}{2004}).
\newblock \bibinfo{title}{Intelligent advisory system for screening mammography}.
\newblock In {\it \bibinfo{booktitle}{IMTC 2004 - Instrumentation and Measurement Technology Conference}\/}.
\newblock \bibinfo{address}{Como, Italy}.
%Type = Article
\bibitem[{Hunyadi(2006)}]{Hunyadi}
\bibinfo{author}{Hunyadi, L.} (\bibinfo{year}{2006}).
\newblock \bibinfo{title}{Grouping, the cognitive basis of recursion in language}.
\newblock {\it \bibinfo{journal}{Argumentum}\/},  {\it \bibinfo{volume}{2}\/}, \bibinfo{pages}{67--114}.
%Type = Inproceedings
\bibitem[{Ioffe \& Szegedy(2015)}]{batchnorm}
\bibinfo{author}{Ioffe, S.}, \& \bibinfo{author}{Szegedy, C.} (\bibinfo{year}{2015}).
\newblock \bibinfo{title}{Batch normalization: Accelerating deep network training by reducing internal covariate shift}.
\newblock In \bibinfo{editor}{F.~Bach}, \& \bibinfo{editor}{D.~Blei} (Eds.), {\it \bibinfo{booktitle}{Proceedings of the 32nd International Conference on Machine Learning}\/} (pp. \bibinfo{pages}{448--456}).
\newblock \bibinfo{address}{Lille, France}: \bibinfo{publisher}{PMLR} volume~\bibinfo{volume}{37} of {\it \bibinfo{series}{Proceedings of Machine Learning Research}\/}.
\newblock \URLprefix \url{https://proceedings.mlr.press/v37/ioffe15.html}.
%Type = Article
\bibitem[{Jelasity \& Dombi(1998)}]{Jelasity}
\bibinfo{author}{Jelasity, M.}, \& \bibinfo{author}{Dombi, J.} (\bibinfo{year}{1998}).
\newblock \bibinfo{title}{Gas, a concept on modeling species in genetic algorithms}.
\newblock {\it \bibinfo{journal}{Artificial Intelligence}\/},  {\it \bibinfo{volume}{99}\/}, \bibinfo{pages}{1--19}.
%Type = Article
\bibitem[{Juh{\'a}sz et~al.(2024)Juh{\'a}sz, Orsz{\'a}g, Balla, Szab{\'o}, Syd{\'o}, Kiss, Csulak, Babity, Dohy, Skoda et~al.}]{juhasz2024blood}
\bibinfo{author}{Juh{\'a}sz, V.}, \bibinfo{author}{Orsz{\'a}g, A.}, \bibinfo{author}{Balla, D.}, \bibinfo{author}{Szab{\'o}, L.}, \bibinfo{author}{Syd{\'o}, N.}, \bibinfo{author}{Kiss, O.}, \bibinfo{author}{Csulak, E.}, \bibinfo{author}{Babity, M.}, \bibinfo{author}{Dohy, Z.}, \bibinfo{author}{Skoda, R.} et~al. (\bibinfo{year}{2024}).
\newblock \bibinfo{title}{Blood test-based age acceleration is inversely associated with high-volume sports activity}.
\newblock {\it \bibinfo{journal}{Medicine and science in sports and exercise}\/},  {\it \bibinfo{volume}{56}\/}, \bibinfo{pages}{868--875}.
%Type = Inproceedings
\bibitem[{Kelen et~al.(2023)Kelen, Petreczky, Kersch \& Bencz{\'u}r}]{kelen2023theoretical}
\bibinfo{author}{Kelen, D.~M.}, \bibinfo{author}{Petreczky, M.}, \bibinfo{author}{Kersch, P.}, \& \bibinfo{author}{Bencz{\'u}r, A.~A.} (\bibinfo{year}{2023}).
\newblock \bibinfo{title}{Theoretical evaluation of asymmetric shapley values for root-cause analysis}.
\newblock In {\it \bibinfo{booktitle}{2023 IEEE International Conference on Data Mining (ICDM)}\/} (pp. \bibinfo{pages}{210--219}).
\newblock \bibinfo{organization}{IEEE}.
%Type = Article
\bibitem[{Kerepesi et~al.(2018)Kerepesi, Dar{\'o}czy, Sturm, Vellai \& Bencz{\'u}r}]{kerepesi2018prediction}
\bibinfo{author}{Kerepesi, C.}, \bibinfo{author}{Dar{\'o}czy, B.}, \bibinfo{author}{Sturm, {\'A}.}, \bibinfo{author}{Vellai, T.}, \& \bibinfo{author}{Bencz{\'u}r, A.} (\bibinfo{year}{2018}).
\newblock \bibinfo{title}{Prediction and characterization of human ageing-related proteins by using machine learning}.
\newblock {\it \bibinfo{journal}{Scientific reports}\/},  {\it \bibinfo{volume}{8}\/}, \bibinfo{pages}{4094}.
%Type = Inproceedings
\bibitem[{Kiss et~al.(2008{\natexlab{a}})Kiss, Varga, Vékony \& Tevesz}]{Kiss1}
\bibinfo{author}{Kiss, D.}, \bibinfo{author}{Varga, D.}, \bibinfo{author}{Vékony, D.}, \& \bibinfo{author}{Tevesz, G.} (\bibinfo{year}{2008}{\natexlab{a}}).
\newblock \bibinfo{title}{Navigation and control systems of an autonomous robot}.
\newblock In {\it \bibinfo{booktitle}{EUROBOT}\/} (pp. \bibinfo{pages}{144--156}).
%Type = Inproceedings
\bibitem[{Kiss et~al.(2008{\natexlab{b}})Kiss, Varga, Vékony \& Tevesz}]{Kiss2}
\bibinfo{author}{Kiss, D.}, \bibinfo{author}{Varga, D.}, \bibinfo{author}{Vékony, D.}, \& \bibinfo{author}{Tevesz, G.} (\bibinfo{year}{2008}{\natexlab{b}}).
\newblock \bibinfo{title}{Navigation and control systems of an autonomous robot}.
\newblock In {\it \bibinfo{booktitle}{EUROBOT}\/} (pp. \bibinfo{pages}{144--156}).
%Type = Inproceedings
\bibitem[{Kocsis \& Szepesvári(2006)}]{Kocsis}
\bibinfo{author}{Kocsis, L.}, \& \bibinfo{author}{Szepesvári, C.} (\bibinfo{year}{2006}).
\newblock \bibinfo{title}{Bandit based monte-carlo planning}.
\newblock In {\it \bibinfo{booktitle}{European Conference on Machine Learning}\/} (pp. \bibinfo{pages}{282--293}).
%Type = Article
\bibitem[{Kocsor \& Toth(2004)}]{Kocsor}
\bibinfo{author}{Kocsor, A.}, \& \bibinfo{author}{Toth, L.} (\bibinfo{year}{2004}).
\newblock \bibinfo{title}{Application of kernel-based feature space transformations and learning methods to phoneme classification}.
\newblock {\it \bibinfo{journal}{Applied Intelligence}\/},  {\it \bibinfo{volume}{21}\/}, \bibinfo{pages}{129--142}.
%Type = Article
\bibitem[{Kov{\'a}cs(2021)}]{nyeki2021application}
\bibinfo{author}{Kov{\'a}cs, A.} (\bibinfo{year}{2021}).
\newblock \bibinfo{title}{Application of spatio-temporal data in site-specific maize yield prediction with machine learning methods}.
\newblock {\it \bibinfo{journal}{Precision Agriculture}\/},  {\it \bibinfo{volume}{22}\/}, \bibinfo{pages}{1397--1415}.
%Type = Article
\bibitem[{Kurent et~al.(2024)Kurent, Friedman, Aloisio, Pasca, Tomasi \& Brank}]{kurent2024bayesian}
\bibinfo{author}{Kurent, B.}, \bibinfo{author}{Friedman, N.}, \bibinfo{author}{Aloisio, A.}, \bibinfo{author}{Pasca, D.}, \bibinfo{author}{Tomasi, R.}, \& \bibinfo{author}{Brank, B.} (\bibinfo{year}{2024}).
\newblock \bibinfo{title}{Bayesian model updating of eight-storey clt building using modal data}.
\newblock {\it \bibinfo{journal}{Probabilistic Engineering Mechanics}\/},  (p. \bibinfo{pages}{103642}).
%Type = Article
\bibitem[{Kóczy \& Hirota(1993{\natexlab{a}})}]{Koczy1}
\bibinfo{author}{Kóczy, L.}, \& \bibinfo{author}{Hirota, K.} (\bibinfo{year}{1993}{\natexlab{a}}).
\newblock \bibinfo{title}{Approximate reasoning by linear rule interpolation and general approximation}.
\newblock {\it \bibinfo{journal}{International Journal of Approximate Reasoning}\/},  {\it \bibinfo{volume}{9}\/}, \bibinfo{pages}{197--225}.
%Type = Article
\bibitem[{Kóczy \& Hirota(1993{\natexlab{b}})}]{Koczy2}
\bibinfo{author}{Kóczy, L.}, \& \bibinfo{author}{Hirota, K.} (\bibinfo{year}{1993}{\natexlab{b}}).
\newblock \bibinfo{title}{Interpolative reasoning with insufficient evidence in sparse fuzzy rule bases}.
\newblock {\it \bibinfo{journal}{Information Sciences}\/},  {\it \bibinfo{volume}{71}\/}, \bibinfo{pages}{169--201}.
%Type = Article
\bibitem[{Lengyel et~al.(2006)Lengyel, Levendovszky, Mezei, Forstner \& Charaf}]{Lengyel}
\bibinfo{author}{Lengyel, L.}, \bibinfo{author}{Levendovszky, T.}, \bibinfo{author}{Mezei, G.}, \bibinfo{author}{Forstner, B.}, \& \bibinfo{author}{Charaf, H.} (\bibinfo{year}{2006}).
\newblock \bibinfo{title}{Metamodel-based model transformation with aspect-oriented constraints}.
\newblock {\it \bibinfo{journal}{Electronic Notes in Theoretical Computer Science}\/},  {\it \bibinfo{volume}{152}\/}, \bibinfo{pages}{111--123}.
%Type = Inproceedings
\bibitem[{Li et~al.(2017)Li, Chang, Cheng, Yang \& Poczos}]{NIPS2017MMDGAN}
\bibinfo{author}{Li, C.-L.}, \bibinfo{author}{Chang, W.-C.}, \bibinfo{author}{Cheng, Y.}, \bibinfo{author}{Yang, Y.}, \& \bibinfo{author}{Poczos, B.} (\bibinfo{year}{2017}).
\newblock \bibinfo{title}{Mmd gan: Towards deeper understanding of moment matching network}.
\newblock In \bibinfo{editor}{I.~Guyon}, \bibinfo{editor}{U.~V. Luxburg}, \bibinfo{editor}{S.~Bengio}, \bibinfo{editor}{H.~Wallach}, \bibinfo{editor}{R.~Fergus}, \bibinfo{editor}{S.~Vishwanathan}, \& \bibinfo{editor}{R.~Garnett} (Eds.), {\it \bibinfo{booktitle}{Advances in Neural Information Processing Systems}\/}.
\newblock \bibinfo{publisher}{Curran Associates, Inc.} volume~\bibinfo{volume}{30}.
\newblock \URLprefix \url{https://proceedings.neurips.cc/paper_files/paper/2017/file/dfd7468ac613286cdbb40872c8ef3b06-Paper.pdf}.
%Type = Article
\bibitem[{Li et~al.(2022)Li, Friedman, Teatini, Benczur, Ye, Zhu \& Zoccarato}]{li2022sensitivity}
\bibinfo{author}{Li, Y.}, \bibinfo{author}{Friedman, N.}, \bibinfo{author}{Teatini, P.}, \bibinfo{author}{Benczur, A.}, \bibinfo{author}{Ye, S.}, \bibinfo{author}{Zhu, L.}, \& \bibinfo{author}{Zoccarato, C.} (\bibinfo{year}{2022}).
\newblock \bibinfo{title}{Sensitivity analysis of factors controlling earth fissures due to excessive groundwater pumping}.
\newblock {\it \bibinfo{journal}{Stochastic Environmental Research and Risk Assessment}\/},  {\it \bibinfo{volume}{36}\/}, \bibinfo{pages}{3911--3928}.
%Type = Inproceedings
\bibitem[{Ligeti-Nagy et~al.(2024)Ligeti-Nagy, Ferenczi, H{\'e}ja, Laki, Vad{\'a}sz, Yang \& V{\'a}radi}]{ligeti2024hulu}
\bibinfo{author}{Ligeti-Nagy, N.}, \bibinfo{author}{Ferenczi, G.}, \bibinfo{author}{H{\'e}ja, E.}, \bibinfo{author}{Laki, L.~J.}, \bibinfo{author}{Vad{\'a}sz, N.}, \bibinfo{author}{Yang, Z.~G.}, \& \bibinfo{author}{V{\'a}radi, T.} (\bibinfo{year}{2024}).
\newblock \bibinfo{title}{Hulu: Hungarian language understanding benchmark kit}.
\newblock In {\it \bibinfo{booktitle}{Proceedings of the 2024 Joint International Conference on Computational Linguistics, Language Resources and Evaluation (LREC-COLING 2024)}\/} (pp. \bibinfo{pages}{8360--8371}).
%Type = Inproceedings
\bibitem[{M{\'a}ndli et~al.(2017)M{\'a}ndli, P{\'a}lovics, Susits \& Bencz{\'u}r}]{mandli2017time}
\bibinfo{author}{M{\'a}ndli, A.}, \bibinfo{author}{P{\'a}lovics, R.}, \bibinfo{author}{Susits, M.}, \& \bibinfo{author}{Bencz{\'u}r, A.~A.} (\bibinfo{year}{2017}).
\newblock \bibinfo{title}{Time series classification for scrap rate prediction in transfer molding}.
\newblock In {\it \bibinfo{booktitle}{3rd SIGKDD Workshop on Mining and Learning from Time Series, Halifax, Nova Scotia, Canada}\/}.
%Type = Inproceedings
\bibitem[{Megyeri et~al.(2020)Megyeri, Hegedűs \& Jelasity}]{Jel1}
\bibinfo{author}{Megyeri, I.}, \bibinfo{author}{Hegedűs, I.}, \& \bibinfo{author}{Jelasity, M.} (\bibinfo{year}{2020}).
\newblock \bibinfo{title}{Adversarial robustness of model sets}.
\newblock In {\it \bibinfo{booktitle}{2020 International Joint Conference on Neural Networks (IJCNN)}\/} (pp. \bibinfo{pages}{1--8}).
\newblock \DOIprefix\doi{10.1109/IJCNN48605.2020.9206656}.
%Type = Article
\bibitem[{Merluzzi et~al.(2023)Merluzzi, Borsos, Rajatheva, Bencz{\'u}r, Farhadi, Yassine, M{\"u}eck, Barmpounakis, Strinati, Dampahalage et~al.}]{merluzzi2023hexa}
\bibinfo{author}{Merluzzi, M.}, \bibinfo{author}{Borsos, T.}, \bibinfo{author}{Rajatheva, N.}, \bibinfo{author}{Bencz{\'u}r, A.~A.}, \bibinfo{author}{Farhadi, H.}, \bibinfo{author}{Yassine, T.}, \bibinfo{author}{M{\"u}eck, M.~D.}, \bibinfo{author}{Barmpounakis, S.}, \bibinfo{author}{Strinati, E.~C.}, \bibinfo{author}{Dampahalage, D.} et~al. (\bibinfo{year}{2023}).
\newblock \bibinfo{title}{The hexa-x project vision on artificial intelligence and machine learning-driven communication and computation co-design for 6g}.
\newblock {\it \bibinfo{journal}{IEEE Access}\/},  {\it \bibinfo{volume}{11}\/}, \bibinfo{pages}{65620--65648}.
%Type = Article
\bibitem[{Monostori(1993)}]{Monostori1}
\bibinfo{author}{Monostori, L.} (\bibinfo{year}{1993}).
\newblock \bibinfo{title}{A step towards intelligent manufacturing: Modeling and monitoring of manufacturing processes through artificial neural networks}.
\newblock {\it \bibinfo{journal}{CIRP Annals – Manufacturing Technology}\/},  {\it \bibinfo{volume}{42}\/}, \bibinfo{pages}{485--488}. \DOIprefix\doi{10.1016/S0007-8506(07)62491-3}.
%Type = Article
\bibitem[{Monostori et~al.(1996)Monostori, Márkus, Brussel \& Westkämpfer}]{Monostori2}
\bibinfo{author}{Monostori, L.}, \bibinfo{author}{Márkus, A.}, \bibinfo{author}{Brussel, H.~V.}, \& \bibinfo{author}{Westkämpfer, E.} (\bibinfo{year}{1996}).
\newblock \bibinfo{title}{Machine learning approaches to manufacturing}.
\newblock {\it \bibinfo{journal}{CIRP Annals}\/},  {\it \bibinfo{volume}{45}\/}, \bibinfo{pages}{675--712}. \DOIprefix\doi{10.1016/S0007-8506(18)30216-6}.
%Type = Article
\bibitem[{Monostori et~al.(2006)Monostori, Váncza \& Kumara}]{Monostori3}
\bibinfo{author}{Monostori, L.}, \bibinfo{author}{Váncza, J.}, \& \bibinfo{author}{Kumara, S.~R.} (\bibinfo{year}{2006}).
\newblock \bibinfo{title}{Agent-based systems for manufacturing}.
\newblock {\it \bibinfo{journal}{CIRP Annals – Manufacturing Technology}\/},  {\it \bibinfo{volume}{55}\/}, \bibinfo{pages}{697--720}. \DOIprefix\doi{10.1016/j.cirp.2006.10.004}.
%Type = Article
\bibitem[{Márkus \& Hatvany(1987)}]{Markus}
\bibinfo{author}{Márkus, A.}, \& \bibinfo{author}{Hatvany, J.} (\bibinfo{year}{1987}).
\newblock \bibinfo{title}{Matching ai tools to engineering requirements}.
\newblock {\it \bibinfo{journal}{CIRP Annals}\/},  {\it \bibinfo{volume}{36}\/}, \bibinfo{pages}{311--315}. \DOIprefix\doi{10.1016/S0007-8506(07)62611-0}.
%Type = Inproceedings
\bibitem[{Mészáros \& Dobrowiecki(1997)}]{Meszaros}
\bibinfo{author}{Mészáros, T.}, \& \bibinfo{author}{Dobrowiecki, T.~P.} (\bibinfo{year}{1997}).
\newblock \bibinfo{title}{Teamwork in intelligent measurement systems}.
\newblock In {\it \bibinfo{booktitle}{Proc. 1997 IEEE Int. Conf. on Intelligent Engineering Systems (INES'97)}\/} (pp. \bibinfo{pages}{221--225}).
%Type = Article
\bibitem[{Nagy et~al.(2021)Nagy, Kov{\'a}cs \& Benedek}]{nagy2021changegan}
\bibinfo{author}{Nagy, B.}, \bibinfo{author}{Kov{\'a}cs, L.}, \& \bibinfo{author}{Benedek, C.} (\bibinfo{year}{2021}).
\newblock \bibinfo{title}{Changegan: A deep network for change detection in coarsely registered point clouds}.
\newblock {\it \bibinfo{journal}{IEEE Robotics and Automation Letters}\/},  {\it \bibinfo{volume}{6}\/}, \bibinfo{pages}{8277--8284}.
%Type = Article
\bibitem[{Nagy et~al.(2023)Nagy, Szabados-Nacsa, F{\"u}l{\"o}p, Nagyn{\'e}, Galata, Farkas, M{\'e}sz{\'a}ros, Nagy \& Marosi}]{nagy2023interpretable}
\bibinfo{author}{Nagy, B.}, \bibinfo{author}{Szabados-Nacsa, {\'A}.}, \bibinfo{author}{F{\"u}l{\"o}p, G.}, \bibinfo{author}{Nagyn{\'e}, A.~T.}, \bibinfo{author}{Galata, D.~L.}, \bibinfo{author}{Farkas, A.}, \bibinfo{author}{M{\'e}sz{\'a}ros, L.~A.}, \bibinfo{author}{Nagy, Z.~K.}, \& \bibinfo{author}{Marosi, G.} (\bibinfo{year}{2023}).
\newblock \bibinfo{title}{Interpretable artificial neural networks for retrospective qbd of pharmaceutical tablet manufacturing based on a pilot-scale developmental dataset}.
\newblock {\it \bibinfo{journal}{International Journal of Pharmaceutics}\/},  {\it \bibinfo{volume}{633}\/}, \bibinfo{pages}{122620}.
%Type = Article
\bibitem[{Neu \& Szepesvari(2007)}]{Neu2007ApprenticeshipLU}
\bibinfo{author}{Neu, G.}, \& \bibinfo{author}{Szepesvari, C.} (\bibinfo{year}{2007}).
\newblock \bibinfo{title}{Apprenticeship learning using inverse reinforcement learning and gradient methods}.
\newblock {\it \bibinfo{journal}{ArXiv}\/},  {\it \bibinfo{volume}{abs/1206.5264}\/}. \URLprefix \url{https://api.semanticscholar.org/CorpusID:9898063}.
%Type = Book
\bibitem[{Németh \& Olaszy(2010)}]{Nemeth}
\bibinfo{editor}{Németh, G.}, \& \bibinfo{editor}{Olaszy, G.} (Eds.) (\bibinfo{year}{2010}).
\newblock {\it \bibinfo{title}{A magyar beszéd}\/}.
\newblock \bibinfo{publisher}{Akadémiai Kiadó}.
\newblock \URLprefix \url{http://speechlab.tmit.bme.hu/downloads/pdf/A-magyar-beszed.pdf} \bibinfo{note}{708 pages}.
%Type = Article
\bibitem[{Olaszy et~al.(2000)Olaszy, Németh, Olaszi et~al.}]{Olaszy}
\bibinfo{author}{Olaszy, G.}, \bibinfo{author}{Németh, G.}, \bibinfo{author}{Olaszi, P.} et~al. (\bibinfo{year}{2000}).
\newblock \bibinfo{title}{Profivox—a hungarian text-to-speech system for telecommunications applications}.
\newblock {\it \bibinfo{journal}{International Journal of Speech Technology}\/},  {\it \bibinfo{volume}{3}\/}, \bibinfo{pages}{201--215}. \DOIprefix\doi{10.1023/A:1026558915015}.
%Type = Inproceedings
\bibitem[{Oravecz et~al.(2014)Oravecz, Váradi \& Sass}]{Oravecz}
\bibinfo{author}{Oravecz, C.}, \bibinfo{author}{Váradi, T.}, \& \bibinfo{author}{Sass, B.} (\bibinfo{year}{2014}).
\newblock \bibinfo{title}{The hungarian gigaword corpus}.
\newblock In \bibinfo{editor}{N.~Calzolari}, \bibinfo{editor}{K.~Choukri}, \bibinfo{editor}{T.~Declerck}, \bibinfo{editor}{H.~Loftsson}, \bibinfo{editor}{B.~Maegaard}, \bibinfo{editor}{J.~Mariani}, \bibinfo{editor}{A.~Moreno}, \bibinfo{editor}{J.~Odijk}, \& \bibinfo{editor}{S.~Piperidis} (Eds.), {\it \bibinfo{booktitle}{LREC 2014 – Ninth International Conference on Language Resources and Evaluation}\/} (pp. \bibinfo{pages}{1719--1723}).
\newblock \bibinfo{address}{Lisbon, Portugal}: \bibinfo{publisher}{European Language Resources Association (ELRA)}.
%Type = Article
\bibitem[{Ormandi et~al.(2013)Ormandi, Hegedus \& Jelasity}]{Ormandi}
\bibinfo{author}{Ormandi, R.}, \bibinfo{author}{Hegedus, I.}, \& \bibinfo{author}{Jelasity, M.} (\bibinfo{year}{2013}).
\newblock \bibinfo{title}{Gossip learning with linear models on fully distributed data}.
\newblock {\it \bibinfo{journal}{Concurrency and Computation: Practice and Experience}\/},  {\it \bibinfo{volume}{25}\/}, \bibinfo{pages}{556--571}. \DOIprefix\doi{10.1002/cpe.2858}.
%Type = Book
\bibitem[{Papp(1969)}]{PappF}
\bibinfo{author}{Papp, F.} (\bibinfo{year}{1969}).
\newblock {\it \bibinfo{title}{A magyar nyelv szóvégmutató szótára}\/}.
\newblock \bibinfo{address}{Budapest}: \bibinfo{publisher}{Akadémiai Kiadó}.
%Type = Inproceedings
\bibitem[{Papp et~al.(1989)Papp, Dobrowiecki, Vadász \& Tilly}]{Papp2}
\bibinfo{author}{Papp, Z.}, \bibinfo{author}{Dobrowiecki, T.}, \bibinfo{author}{Vadász, B.}, \& \bibinfo{author}{Tilly, K.} (\bibinfo{year}{1989}).
\newblock \bibinfo{title}{Expert system architecture for real-time process supervisor applications}.
\newblock In \bibinfo{editor}{G.~Rzevski} (Ed.), {\it \bibinfo{booktitle}{Proc. of the 4th Int. Conf. on the Applications of Artificial Intelligence in Engineering}\/} (pp. \bibinfo{pages}{223--240}).
\newblock \bibinfo{address}{Cambridge, UK}: \bibinfo{publisher}{Springer-Verlag}.
%Type = Article
\bibitem[{Papp et~al.(1988)Papp, Peceli, Bago \& Pataki}]{Papp1}
\bibinfo{author}{Papp, Z.}, \bibinfo{author}{Peceli, G.}, \bibinfo{author}{Bago, B.}, \& \bibinfo{author}{Pataki, B.} (\bibinfo{year}{1988}).
\newblock \bibinfo{title}{Intelligent medical instruments}.
\newblock {\it \bibinfo{journal}{IEEE Engineering in Medicine and Biology Magazine}\/},  (pp. \bibinfo{pages}{18--23}).
%Type = Article
\bibitem[{Paragh et~al.(2018)Paragh, Harangi, Kar{\'a}nyi, Dar{\'o}czy, N{\'e}meth \& F{\"u}l{\"o}p}]{paragh2018identifying}
\bibinfo{author}{Paragh, G.}, \bibinfo{author}{Harangi, M.}, \bibinfo{author}{Kar{\'a}nyi, Z.}, \bibinfo{author}{Dar{\'o}czy, B.}, \bibinfo{author}{N{\'e}meth, {\'A}.}, \& \bibinfo{author}{F{\"u}l{\"o}p, P.} (\bibinfo{year}{2018}).
\newblock \bibinfo{title}{Identifying patients with familial hypercholesterolemia using data mining methods in the northern great plain region of hungary}.
\newblock {\it \bibinfo{journal}{Atherosclerosis}\/},  {\it \bibinfo{volume}{277}\/}, \bibinfo{pages}{262--266}.
%Type = Article
\bibitem[{Pethő et~al.(1999)Pethő, Zimmer, Gebel \& Herrmann}]{Pethő}
\bibinfo{author}{Pethő, A.}, \bibinfo{author}{Zimmer, H.}, \bibinfo{author}{Gebel, J.}, \& \bibinfo{author}{Herrmann, E.} (\bibinfo{year}{1999}).
\newblock \bibinfo{title}{Computing all s-integral points on elliptic curves}.
\newblock {\it \bibinfo{journal}{Mathematical Proceedings of the Cambridge Philosophical Society}\/},  {\it \bibinfo{volume}{127}\/}, \bibinfo{pages}{383--402}.
%Type = Article
\bibitem[{P{\'o}sfai et~al.(2024)P{\'o}sfai, Szegedy, Ba{\v{c}}i{\'c}, Blagojevi{\'c}, Ab{\'e}rt, Kert{\'e}sz, Lov{\'a}sz \& Barab{\'a}si}]{posfai2024impact}
\bibinfo{author}{P{\'o}sfai, M.}, \bibinfo{author}{Szegedy, B.}, \bibinfo{author}{Ba{\v{c}}i{\'c}, I.}, \bibinfo{author}{Blagojevi{\'c}, L.}, \bibinfo{author}{Ab{\'e}rt, M.}, \bibinfo{author}{Kert{\'e}sz, J.}, \bibinfo{author}{Lov{\'a}sz, L.}, \& \bibinfo{author}{Barab{\'a}si, A.-L.} (\bibinfo{year}{2024}).
\newblock \bibinfo{title}{Impact of physicality on network structure}.
\newblock {\it \bibinfo{journal}{Nature Physics}\/},  {\it \bibinfo{volume}{20}\/}, \bibinfo{pages}{142--149}.
%Type = Inproceedings
\bibitem[{Prószéky(1995)}]{Prószéky1}
\bibinfo{author}{Prószéky, G.} (\bibinfo{year}{1995}).
\newblock \bibinfo{title}{Humor – a morphological system for corpus analysis}.
\newblock In \bibinfo{editor}{H.~Rettig}, \bibinfo{editor}{J.~Pajzs}, \& \bibinfo{editor}{G.~Kiss} (Eds.), {\it \bibinfo{booktitle}{Language Resources for Language Technology (Proceedings of the 1st TELRI Seminar)}\/} (pp. \bibinfo{pages}{149--158}).
\newblock \bibinfo{address}{Tihany}.
%Type = Incollection
\bibitem[{Prószéky(2000)}]{Prószéky2}
\bibinfo{author}{Prószéky, G.} (\bibinfo{year}{2000}).
\newblock \bibinfo{title}{A magyar morfológia számítógépes kezelése}.
\newblock In \bibinfo{editor}{F.~Kiefer} (Ed.), {\it \bibinfo{booktitle}{Morfológia (Strukturális magyar nyelvtan 3)}\/} (pp. \bibinfo{pages}{1024--1065}).
\newblock \bibinfo{address}{Budapest}: \bibinfo{publisher}{Akadémiai Kiadó}.
%Type = Inproceedings
\bibitem[{Prószéky \& Tihanyi(1993)}]{Prószéky3}
\bibinfo{author}{Prószéky, G.}, \& \bibinfo{author}{Tihanyi, L.} (\bibinfo{year}{1993}).
\newblock \bibinfo{title}{Helyette – inflectional thesaurus for agglutinative languages}.
\newblock In {\it \bibinfo{booktitle}{Proceedings of the 6th Conference of the European Chapter of the Association for Computational Linguistics}\/} (p. \bibinfo{pages}{473}).
\newblock \bibinfo{address}{Utrecht, The Netherlands}.
%Type = Article
\bibitem[{Ribli et~al.(2018)Ribli, Horv{\'a}th, Unger, Pollner \& Csabai}]{ribli2018detecting}
\bibinfo{author}{Ribli, D.}, \bibinfo{author}{Horv{\'a}th, A.}, \bibinfo{author}{Unger, Z.}, \bibinfo{author}{Pollner, P.}, \& \bibinfo{author}{Csabai, I.} (\bibinfo{year}{2018}).
\newblock \bibinfo{title}{Detecting and classifying lesions in mammograms with deep learning}.
\newblock {\it \bibinfo{journal}{Scientific reports}\/},  {\it \bibinfo{volume}{8}\/}, \bibinfo{pages}{4165}.
%Type = Article
\bibitem[{Roska \& Chua(1993)}]{Roska}
\bibinfo{author}{Roska, T.}, \& \bibinfo{author}{Chua, L.~O.} (\bibinfo{year}{1993}).
\newblock \bibinfo{title}{The cnn universal machine: An analogic array computer}.
\newblock {\it \bibinfo{journal}{IEEE Transactions on Circuits and Systems II - Analog and Digital Signal Processing}\/},  {\it \bibinfo{volume}{40}\/}, \bibinfo{pages}{163--173}.
%Type = Inproceedings
\bibitem[{Rozemberczki et~al.(2021)Rozemberczki, Scherer, He, Panagopoulos, Riedel, Astefanoaei, Kiss, Beres, Lopez, Collignon et~al.}]{rozemberczki2021pytorch}
\bibinfo{author}{Rozemberczki, B.}, \bibinfo{author}{Scherer, P.}, \bibinfo{author}{He, Y.}, \bibinfo{author}{Panagopoulos, G.}, \bibinfo{author}{Riedel, A.}, \bibinfo{author}{Astefanoaei, M.}, \bibinfo{author}{Kiss, O.}, \bibinfo{author}{Beres, F.}, \bibinfo{author}{Lopez, G.}, \bibinfo{author}{Collignon, N.} et~al. (\bibinfo{year}{2021}).
\newblock \bibinfo{title}{Pytorch geometric temporal: Spatiotemporal signal processing with neural machine learning models}.
\newblock In {\it \bibinfo{booktitle}{Proceedings of the 30th ACM international conference on information \& knowledge management}\/} (pp. \bibinfo{pages}{4564--4573}).
%Type = Article
\bibitem[{Seb{\H{o}}k \& Kacsuk(2021)}]{sebok2021multiclass}
\bibinfo{author}{Seb{\H{o}}k, M.}, \& \bibinfo{author}{Kacsuk, Z.} (\bibinfo{year}{2021}).
\newblock \bibinfo{title}{The multiclass classification of newspaper articles with machine learning: The hybrid binary snowball approach}.
\newblock {\it \bibinfo{journal}{Political Analysis}\/},  {\it \bibinfo{volume}{29}\/}, \bibinfo{pages}{236--249}.
%Type = Inproceedings
\bibitem[{Senellart et~al.(2003)Senellart, Dienes \& Váradi}]{Senellart}
\bibinfo{author}{Senellart, J.}, \bibinfo{author}{Dienes, P.}, \& \bibinfo{author}{Váradi, T.} (\bibinfo{year}{2003}).
\newblock \bibinfo{title}{New generation systran translation system}.
\newblock In \bibinfo{editor}{J.~Senellart}, \bibinfo{editor}{J.~Yang}, \& \bibinfo{editor}{A.~Rebollo} (Eds.), {\it \bibinfo{booktitle}{Proceedings of MT Summit IX}\/}.
%Type = Article
\bibitem[{Shaheen et~al.(2023)Shaheen, Kocsis \& N{\'e}meth}]{shaheen2023data}
\bibinfo{author}{Shaheen, B.}, \bibinfo{author}{Kocsis, {\'A}.}, \& \bibinfo{author}{N{\'e}meth, I.} (\bibinfo{year}{2023}).
\newblock \bibinfo{title}{Data-driven failure prediction and rul estimation of mechanical components using accumulative artificial neural networks}.
\newblock {\it \bibinfo{journal}{Engineering Applications of Artificial Intelligence}\/},  {\it \bibinfo{volume}{119}\/}, \bibinfo{pages}{105749}.
%Type = Inproceedings
\bibitem[{Siegler \& Vámos(1981)}]{Siegler}
\bibinfo{author}{Siegler, A.}, \& \bibinfo{author}{Vámos, T.} (\bibinfo{year}{1981}).
\newblock \bibinfo{title}{Intelligent robot action planning}.
\newblock In \bibinfo{editor}{H.~Akashi} (Ed.), {\it \bibinfo{booktitle}{Control Science and Technology for the Progress of Society, Proceedings of the 8th Triennial World Congress of the International Federation of Automatic Control}\/} (pp. \bibinfo{pages}{103--108}).
\newblock \bibinfo{address}{Oxford, United Kingdom}: \bibinfo{publisher}{Pergamon Press}.
%Type = Article
\bibitem[{Szabó(2015)}]{Szabo}
\bibinfo{author}{Szabó, P.~G.} (\bibinfo{year}{2015}).
\newblock \bibinfo{title}{Kalmár {László}, a számítástudomány hazai úttörője}.
\newblock {\it \bibinfo{journal}{Alkalmazott Matematikai Lapok}\/},  {\it \bibinfo{volume}{32}\/}, \bibinfo{pages}{79--94}.
%Type = Misc
\bibitem[{Szegedy et~al.(2014)Szegedy, Zaremba, Sutskever, Bruna, Erhan, Goodfellow \& Fergus}]{szegedy2014}
\bibinfo{author}{Szegedy, C.}, \bibinfo{author}{Zaremba, W.}, \bibinfo{author}{Sutskever, I.}, \bibinfo{author}{Bruna, J.}, \bibinfo{author}{Erhan, D.}, \bibinfo{author}{Goodfellow, I.}, \& \bibinfo{author}{Fergus, R.} (\bibinfo{year}{2014}).
\newblock \bibinfo{title}{Intriguing properties of neural networks}.
\newblock \URLprefix \url{https://arxiv.org/abs/1312.6199}. \href{http://arxiv.org/abs/1312.6199}{\tt arXiv:1312.6199}.
%Type = Article
\bibitem[{Szemenyei \& Estivill-Castro(2022)}]{szemenyei2022fully}
\bibinfo{author}{Szemenyei, M.}, \& \bibinfo{author}{Estivill-Castro, V.} (\bibinfo{year}{2022}).
\newblock \bibinfo{title}{Fully neural object detection solutions for robot soccer}.
\newblock {\it \bibinfo{journal}{Neural Computing and Applications}\/},  {\it \bibinfo{volume}{34}\/}, \bibinfo{pages}{21419--21432}.
%Type = Article
\bibitem[{Szirányi \& Csicsvári(1993)}]{Sziranyi}
\bibinfo{author}{Szirányi, T.}, \& \bibinfo{author}{Csicsvári, J.} (\bibinfo{year}{1993}).
\newblock \bibinfo{title}{High speed character recognition using a dual cellular neural network architecture}.
\newblock {\it \bibinfo{journal}{IEEE Transactions on Circuits and Systems II - Analog and Digital Signal Processing}\/},  {\it \bibinfo{volume}{40}\/}, \bibinfo{pages}{223--231}.
%Type = Inproceedings
\bibitem[{Szita \& Lőrincz(2009)}]{Szita}
\bibinfo{author}{Szita, I.}, \& \bibinfo{author}{Lőrincz, A.} (\bibinfo{year}{2009}).
\newblock \bibinfo{title}{Optimistic initialization and greediness lead to polynomial time learning in factored mdps}.
\newblock In {\it \bibinfo{booktitle}{Proceedings of the 26th Annual International Conference on Machine Learning}\/} (pp. \bibinfo{pages}{1001--1008}).
%Type = Article
\bibitem[{Sztipanovits(1989)}]{Sztipanovits}
\bibinfo{author}{Sztipanovits, J.} (\bibinfo{year}{1989}).
\newblock \bibinfo{title}{Intelligent instruments}.
\newblock {\it \bibinfo{journal}{Measurement}\/},  {\it \bibinfo{volume}{7}\/}.
%Type = Misc
\bibitem[{Sántáné-Tóth(1996)}]{SantaneToth2}
\bibinfo{author}{Sántáné-Tóth, E.} (\bibinfo{year}{1996}).
\newblock \bibinfo{title}{Magyar mesterséges intelligencia bibliográfia}.
\newblock \bibinfo{howpublished}{\url{https://mek.oszk.hu/00000/00024/00024.htm}}.
%Type = Misc
\bibitem[{Sántáné-Tóth(2007)}]{SantaneToth1}
\bibinfo{author}{Sántáné-Tóth, E.} (\bibinfo{year}{2007}).
\newblock \bibinfo{title}{Artificial intelligence in hungary – the first 20 years}.
\newblock \bibinfo{howpublished}{\url{https://itf.njszt.hu/wp-content/uploads/AI-in-H.pdf}}.
%Type = Article
\bibitem[{Tak{\'a}cs et~al.(2009)Tak{\'a}cs, Pil{\'a}szy, N{\'e}meth \& Tikk}]{takacs2009scalable}
\bibinfo{author}{Tak{\'a}cs, G.}, \bibinfo{author}{Pil{\'a}szy, I.}, \bibinfo{author}{N{\'e}meth, B.}, \& \bibinfo{author}{Tikk, D.} (\bibinfo{year}{2009}).
\newblock \bibinfo{title}{Scalable collaborative filtering approaches for large recommender systems}.
\newblock {\it \bibinfo{journal}{The Journal of Machine Learning Research}\/},  {\it \bibinfo{volume}{10}\/}, \bibinfo{pages}{623--656}.
%Type = Inproceedings
\bibitem[{Toth et~al.(2008)Toth, Frankel, Gosztolya \& King}]{Toth2}
\bibinfo{author}{Toth, L.}, \bibinfo{author}{Frankel, J.}, \bibinfo{author}{Gosztolya, G.}, \& \bibinfo{author}{King, S.} (\bibinfo{year}{2008}).
\newblock \bibinfo{title}{Cross-lingual portability of mlp-based tandem features - a case study for english and hungarian}.
\newblock In {\it \bibinfo{booktitle}{Proceedings of Interspeech 2008}\/} (pp. \bibinfo{pages}{2695--2698}).
%Type = Article
\bibitem[{Toth \& Kocsor(2007)}]{Toth1}
\bibinfo{author}{Toth, L.}, \& \bibinfo{author}{Kocsor, A.} (\bibinfo{year}{2007}).
\newblock \bibinfo{title}{A segment-based interpretation of hmm/ann hybrids}.
\newblock {\it \bibinfo{journal}{Computer Speech And Language}\/},  {\it \bibinfo{volume}{21}\/}, \bibinfo{pages}{562--578}.
%Type = Inproceedings
\bibitem[{Valyon \& Horvath(2001)}]{Valyon}
\bibinfo{author}{Valyon, J.}, \& \bibinfo{author}{Horvath, G.} (\bibinfo{year}{2001}).
\newblock \bibinfo{title}{A hybrid intelligent system for image matching, used as preprocessing for signature verification}.
\newblock In {\it \bibinfo{booktitle}{Proc. Artificial Neural Nets and Genetic Algorithms Conference}\/} (pp. \bibinfo{pages}{367--370}).
%Type = Inproceedings
\bibitem[{Varga et~al.(2003)}]{Varga}
\bibinfo{author}{Varga, P.} et~al. (\bibinfo{year}{2003}).
\newblock \bibinfo{title}{An ontology based information retrieval system}.
\newblock In {\it \bibinfo{booktitle}{Proc. of the IEA/AIE-2003}\/}.
\newblock \bibinfo{address}{Loughborough, UK}: \bibinfo{publisher}{Springer-Verlag} volume \bibinfo{volume}{2718} of {\it \bibinfo{series}{Lecture Notes in Computer Science}\/}.
%Type = Article
\bibitem[{Viharos et~al.(2021)Viharos, Kis, Fodor \& B{\"u}ki}]{viharos2021adaptive}
\bibinfo{author}{Viharos, Z.~J.}, \bibinfo{author}{Kis, K.~B.}, \bibinfo{author}{Fodor, {\'A}.}, \& \bibinfo{author}{B{\"u}ki, M.~I.} (\bibinfo{year}{2021}).
\newblock \bibinfo{title}{Adaptive, hybrid feature selection (ahfs)}.
\newblock {\it \bibinfo{journal}{Pattern Recognition}\/},  {\it \bibinfo{volume}{116}\/}, \bibinfo{pages}{107932}.
%Type = Incollection
\bibitem[{Vámos(1977)}]{Vámos1}
\bibinfo{author}{Vámos, T.} (\bibinfo{year}{1977}).
\newblock \bibinfo{title}{Industrial objects and machine parts recognition}.
\newblock In \bibinfo{editor}{K.~Fu} (Ed.), {\it \bibinfo{booktitle}{Applications of Syntactic Pattern Recognition}\/} (pp. \bibinfo{pages}{243--267}).
\newblock \bibinfo{address}{Berlin, Germany, Heidelberg, Germany}: \bibinfo{publisher}{Springer-Verlag}.
%Type = Book
\bibitem[{Vámos(1991)}]{Vámos2}
\bibinfo{author}{Vámos, T.} (\bibinfo{year}{1991}).
\newblock {\it \bibinfo{title}{Computer Epistemology: A Treatise on the Feasibility of the Unfeasible or Old Ideas Brewed New}\/}.
\newblock \bibinfo{address}{Singapore}: \bibinfo{publisher}{World Scientific}.
%Type = Article
\bibitem[{Váncza \& Márkus(1991)}]{Vancza}
\bibinfo{author}{Váncza, J.}, \& \bibinfo{author}{Márkus, A.} (\bibinfo{year}{1991}).
\newblock \bibinfo{title}{Genetic algorithms in process planning}.
\newblock {\it \bibinfo{journal}{Computers in Industry}\/},  {\it \bibinfo{volume}{17}\/}, \bibinfo{pages}{181--194}. \DOIprefix\doi{10.1016/0166-3615(91)90031-4}.
%Type = Article
\bibitem[{Váradi \& Kiss(2001)}]{Váradi}
\bibinfo{author}{Váradi, T.}, \& \bibinfo{author}{Kiss, G.} (\bibinfo{year}{2001}).
\newblock \bibinfo{title}{Equivalence and non-equivalence in parallel corpora}.
\newblock {\it \bibinfo{journal}{International Journal of Corpus Linguistics}\/},  {\it \bibinfo{volume}{6}\/}, \bibinfo{pages}{167--177}.
\newblock \bibinfo{note}{Special Issue}.
%Type = Inproceedings
\bibitem[{Várkonyi-Kóczy et~al.(1998)Várkonyi-Kóczy, Péceli, Dobrowiecki \& Kovácsházy}]{VarkonyiKoczy}
\bibinfo{author}{Várkonyi-Kóczy, A.~R.}, \bibinfo{author}{Péceli, G.}, \bibinfo{author}{Dobrowiecki, T.~P.}, \& \bibinfo{author}{Kovácsházy, T.} (\bibinfo{year}{1998}).
\newblock \bibinfo{title}{Iterative fuzzy model inversion}.
\newblock In {\it \bibinfo{booktitle}{The 1998 IEEE Int. Conf. on Fuzzy Systems, IEEE World Congress on Computational Intelligence}\/}.
%Type = Article
\bibitem[{Zsedrovits et~al.(2016)Zsedrovits, Peter, Bauer, Pencz, Hiba, Gozse, Kisantal, Nemeth, Nagy, Vanek et~al.}]{zsedrovits2016onboard}
\bibinfo{author}{Zsedrovits, T.}, \bibinfo{author}{Peter, P.}, \bibinfo{author}{Bauer, P.}, \bibinfo{author}{Pencz, B. J.~M.}, \bibinfo{author}{Hiba, A.}, \bibinfo{author}{Gozse, I.}, \bibinfo{author}{Kisantal, M.}, \bibinfo{author}{Nemeth, M.}, \bibinfo{author}{Nagy, Z.}, \bibinfo{author}{Vanek, B.} et~al. (\bibinfo{year}{2016}).
\newblock \bibinfo{title}{Onboard visual sense and avoid system for small aircraft}.
\newblock {\it \bibinfo{journal}{IEEE Aerospace and Electronic Systems Magazine}\/},  {\it \bibinfo{volume}{31}\/}, \bibinfo{pages}{18--27}.

\end{thebibliography}

\end{document}